\documentclass[12pt, draftcls, onecolumn]{IEEEtran}
\usepackage{cite,amsthm}
\IEEEoverridecommandlockouts
\theoremstyle{definition}
\newtheorem{theorem}{Theorem}
\newtheorem{definition}{Definition}
\newtheorem{lemma}{Lemma}

\usepackage{setspace,multirow}
\usepackage{amsmath,amssymb,amsfonts}
\usepackage{algorithm}  
\usepackage{algorithmic}
\usepackage{subfigure}
\usepackage{graphicx}
\usepackage{textcomp}
\def\BibTeX{{\rm B\kern-.05em{\sc i\kern-.025em b}\kern-.08em
    T\kern-.1667em\lower.7ex\hbox{E}\kern-.125emX}}

\makeatletter
\renewcommand\normalsize{
	\abovedisplayskip 3\p@ \@plus5\p@ \@minus7\p@
	\belowdisplayskip \abovedisplayskip
	\let\@listi\@listI}
\makeatother

\begin{document}

\title{Real-Time Reconstruction of Counting Process through Queues}

\author{Meng~Wang, Wei~Chen,~\IEEEmembership{Senior Member, IEEE,} Anthony~Ephremides,~\IEEEmembership{Life~Fellow,~IEEE}
	
\thanks{M. Wang and W. Chen are with the Department of Electrical Engineering and Beijing National Research Center for Information Science and Technology, Tsinghua University, Beijing 100084, China (email: m-wang14@mails.tsinghua.edu.cn and wchen@tsinghua.edu.cn).}

\thanks{A. Ephremides is with the Department of Electrical and Computer Engineering and the Institute for System Research, University of Maryland, College Park, MD 20740, USA (email: etony@umd.edu).}

\thanks{This paper was submitted in part to the IEEE International Conference on Communications (ICC) 2019.}

\thanks{The work of Meng Wang and Wei Chen is supported by the National Science Foundation of China under Grant No. 61671269 and the National Program for Special Support for Eminent Professionals (10000-Talent Program). The work of Anthony Ephremides is supported by the U.S. Office of Naval Research under Grant N000141812046 and the U.S. National Science Foundation under Grants CCF1420651, CNS1526309, and CCFR1813078. }
}

\maketitle

\begin{abstract}
For the emerging Internet of Things (IoT), one of the most critical problems is the real-time reconstruction of signals from a set of aged measurements. During the reconstruction, distortion occurs between the observed signal and the reconstructed signal due to sampling and transmission. In this paper, we focus on minimizing the average distortion defined as the $1$-$norm$ of the difference of the two signals under the scenario that a Poisson counting process is reconstructed in real-time on a remote monitor. Especially, we consider the reconstruction under uniform sampling policy and two non-uniform sampling policies, i.e., the threshold-based policy and the zero-wait policy. For each of the policy, we derive the closed-form expression of the average distortion by dividing the overall distortion area into polygons and analyzing their  structures. It turns out that the polygons are built up by sub-polygons that account for distortions caused by sampling and transmission. The closed-form expressions of the average distortion help us find the optimal sampling parameters that achieve the minimum distortion. Simulation results are provided to validate our conclusion.
\end{abstract}


\begin{spacing}{1.6}
	
\section{Introduction}
\vspace{-0mm}
Recently, the new emerging Internet of Things (IoT) has gained massive attraction in various spheres. It is common to reconstruct on-going signals (or functions of the signals) remotely in the IoT scenarios\cite{diaz2016state,7845391}. However, the reconstruction is usually based on the measurements that are under sampled and delayed in the network\cite{li2013compressed}. The under sampled measurements omit the detailed information of the original process and the delayed measurements against the requirement of real-time. Thus, it is an important and interesting work to study how to improve the performance of the reconstruction from a set of under-sampled and aged measurements. Solving this problem brings more challenges to the traditional sampling and signal \mbox{processing techniques}. 

Specifically, on one hand, adopting the Nyquist rate when sampling is undesirable as it will cause crowded traffic in the network and/or sometimes impossible when the signal of interest is analog and without a clear beginning and end \cite{4472240}. On the other hand, in traditional non-causal reconstruction scheme, one needs the signal's samples from the entire time horizon, both in the past and in the future. Perfect reconstruction is achievable but with significant delay spent on waiting for future samples to be taken and delivered \cite{nyquist1928certain, 1447892}. To meet the requirement of real-time, it is more suitable to adopt the causal reconstruction scheme, where one can use the available past samples to estimate and predict the current signal value and the goal is to minimize the reconstruction error. Due to the above mentioned reasons, it is of great importance to study the real-time signal reconstruction with sub-Nyquist rate. 

Many works have been done in this area. On one hand, some of them are interested in the compressive sampling (CS) which is a recently evolved notion\cite{candes2006robust,eldar2015sampling, marvasti2012nonuniform}. By recognizing that the intrinsic information of an analog signal is not solely dictated by its bandwidth, CS provides ways to sample a large class of signals at rates which are far below the Nyquist rate and then reconstruct them asymptotically and exactly via efficient numerical methods \cite{vaswani2016recursive,tropp2007signal,maleki2010optimally}. In \cite{5419067} and \cite{6827963}, the authors proposed a suite of dynamic updating algorithms for solving the $\ell_{1}$-norm minimization problems for the case where the central processing unit receives a continuous stream of measurements (or samples) acquired at a fixed rate. In \cite{wijewardhana2017bayesian}, reconstruction algorithm based on sliding window processing was proposed to improve the performance of the signal recovery. The proposed methods in \cite{5419067,6827963,wijewardhana2017bayesian} provide state-of-the-art performance for the case where central processing unit receives a continuous stream of samples acquired at a fixed rate. However, a standard CS approach often assumes that the signal of interest is a vector of finite length (i.e., already digitized) and reconstructs it based on the previously acquired samples (i.e., the non-causality assumption in classic sampling theory \cite{eldar2015sampling}). Thus, methods that release these constraints need to be further explored.

On the other hand, some works have been done by focusing on deriving the age-aware sampling and reconstruction techniques \cite{sun2017remote,bedewy2016optimizing}. The age of information (AoI) is a topic first proposed in \cite{6195689} and further studied in \cite{6620189,7415972,kam2016controlling} to describe the freshness of the information. To perform causal reconstruction, one commonly used method is to introduce a delay in the reconstruction signal and truncate the remaining non-causal part. Thus, the age of the received samples has an impact on the reconstruction quality of the signal. In \cite{sun2017remote} and \cite{bedewy2016optimizing}, the authors minimized the reconstruction error by solving the instances of the causal sampling problem, and provided better performance than the classic uniform sampling approach. In these works, the defined AoI is a process-independent metric. However, the freshness of the information should depend on the context, including the correlation structure, regular pattern, and some parameter values of the signal, which varies from one process to another. Thus, the original defined AoI needs modification to better measure the freshness of information and play an important role in \mbox{signal reconstruction}. 

In this work, we focus on the remote and real-time process reconstruction problem with a new defined age-related metric, i.e., the average distortion, being the measurement criteria. To understand and make progress in the difficult issue of studying the age-related component of distortion (which is new and unstudied so far), the Poisson counting process is a good first choice since it is relatively simple, tractable, and displays all the aspects of distortion that age induces. Thus, we consider such a system that measurements of an on-going Poisson counting process are sampled and transmitted to a remote monitor, where a new process is reconstructed based on the received aged measurements as an estimation of the original process. We adopt the average distortion defined as the average distance between the original process and the reconstructed process as the metric to evaluate the reconstruction error. In this case, the distortion happens due to sampling and transmission. On one hand, sampled measurements ignore the evolution of the process of interest during the sampling interval and just provide snapshots at the sampling instants. On the other hand, the measurements cannot be received by the remote receiver immediately after being generated due to the transmission which causes inevitable delay. These two kinds of distortion happen simultaneously, twist with each other, and build up the total reconstruction distortion. We study the average distortion under the uniform sampling policy and two non-uniform sampling polices, and try to find the optimal sampling parameters that achieve the minimum average distortion. To this end, the closed-form expressions of the average distortion are built as functions of the sampling parameters. Simulation results validate the correctness of these distortion functions. The contributions of this work are summarized as follows.

\subsubsection{The definition of the average distortion} 
The defined average distortion covers the sampling distortion and the transmission distortion, that occur simultaneously and twist with each other during the real-time reconstruction. At the mean time. it is a process-dependent metric which allows it to better measure the freshness of the received samples compared to the AoI metric.

\subsubsection{The closed-form expressions of the average distortion under different sampling policies} For uniform sampling, we build the average distortion as a function of the sampling rate which is the only key parameter for the uniform sampling policy. In this way, the optimal sampling rate can be found to minimize the average distortion. What's more, an interpolation algorithm is provided to further decrease the average distortion. For the non-uniform sampling policy, we consider the threshold-based policy and the zero-wait policy. The closed-form expressions of the average distortion are derived for both cases.

The paper is organized as follows. Section II describes the system model and gives the definition of the average distortion. Section III focuses on minimizing the average distortion under the uniform sampling method. Section IV studies the average distortion under two non-uniform sampling methods. Simulation results are given in Section V to validate the theoretical results. Finally, Section VI concludes this work and talks about the future work.

\vspace{-0mm}
\section{System Model}
\vspace{-0mm}

\begin{figure}[t]
	\centering
	\includegraphics[width=0.99\columnwidth]{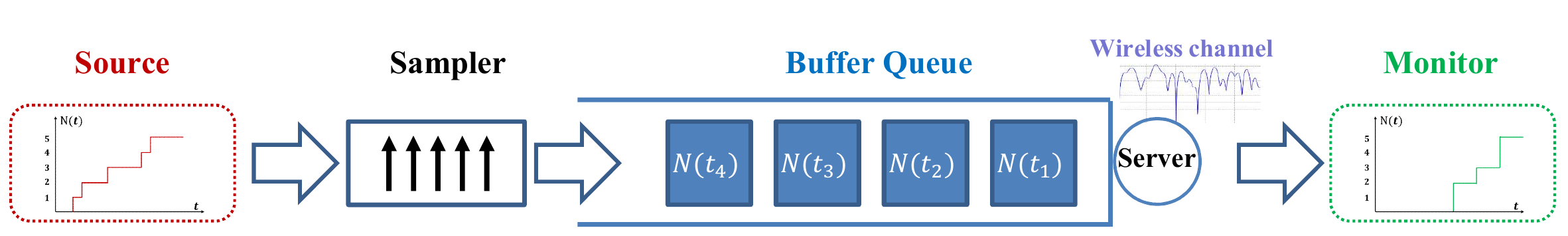}
	\setlength{\abovecaptionskip}{-2mm}
	\caption{System model}
	\label{10151546} 
	\vspace{-0.5cm}
\end{figure}

In this paper, we consider the scenario that a counting process is reconstructed on a remote receiver in real-time  as shown in \figurename\ \ref{10151546}. In detail, there is a counting process $N(t)$ evolving with time on the source node. In order to reconstruct it on a remote monitor, samples of $N(t)$ are obtained by a sampler and then enter a queue waiting to be transmitted. With the received samples, the monitor reconstructs a process $\hat{N}(t)$ as an approximation of the \mbox{original process $N(t)$}. 

\begin{figure*}[t]
	\centering
	\subfigure[The oringinal counting process]{
		\vspace{-0mm}
		\label{model:subfig:b} 
		\includegraphics[width=0.3\columnwidth]{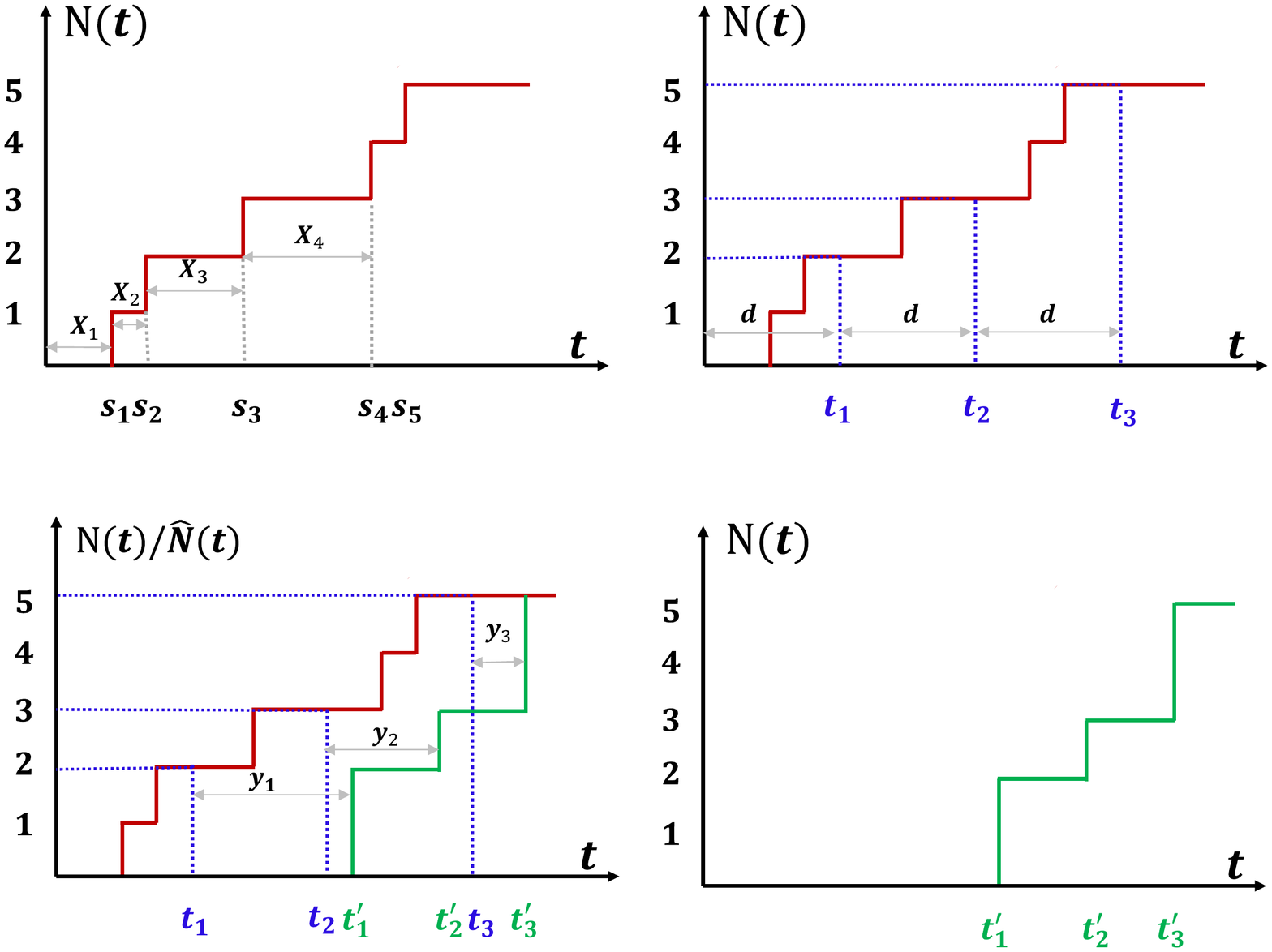}}
	\subfigure[Uniform sampling]{
		\vspace{-0mm}
		\label{model:subfig:c} 
		\includegraphics[width=0.3\columnwidth]{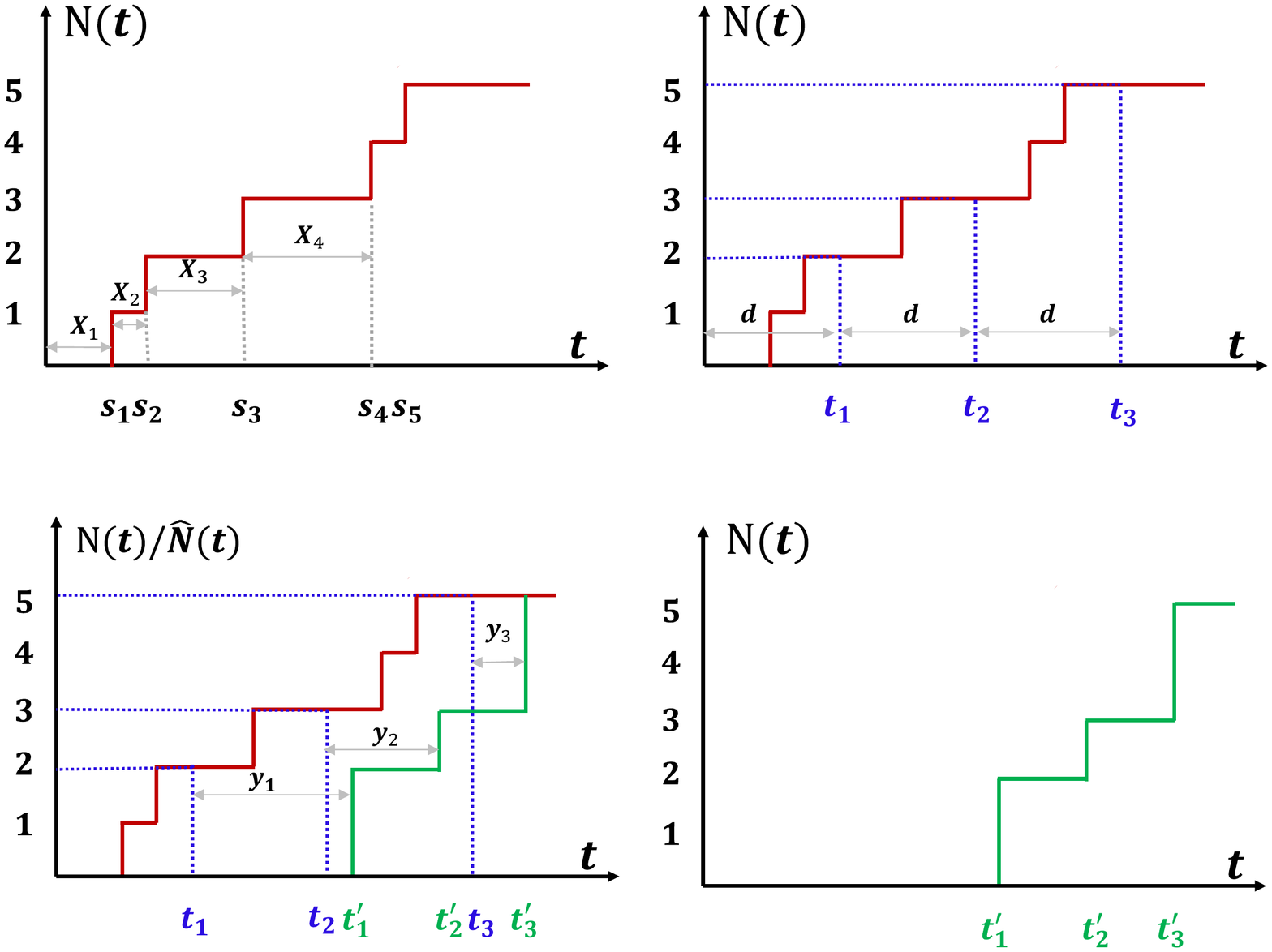}}	
	\subfigure[The reconstructed process]{
		\vspace{-00mm}
		\label{model:subfig:d} 
		\includegraphics[width=0.3\columnwidth]{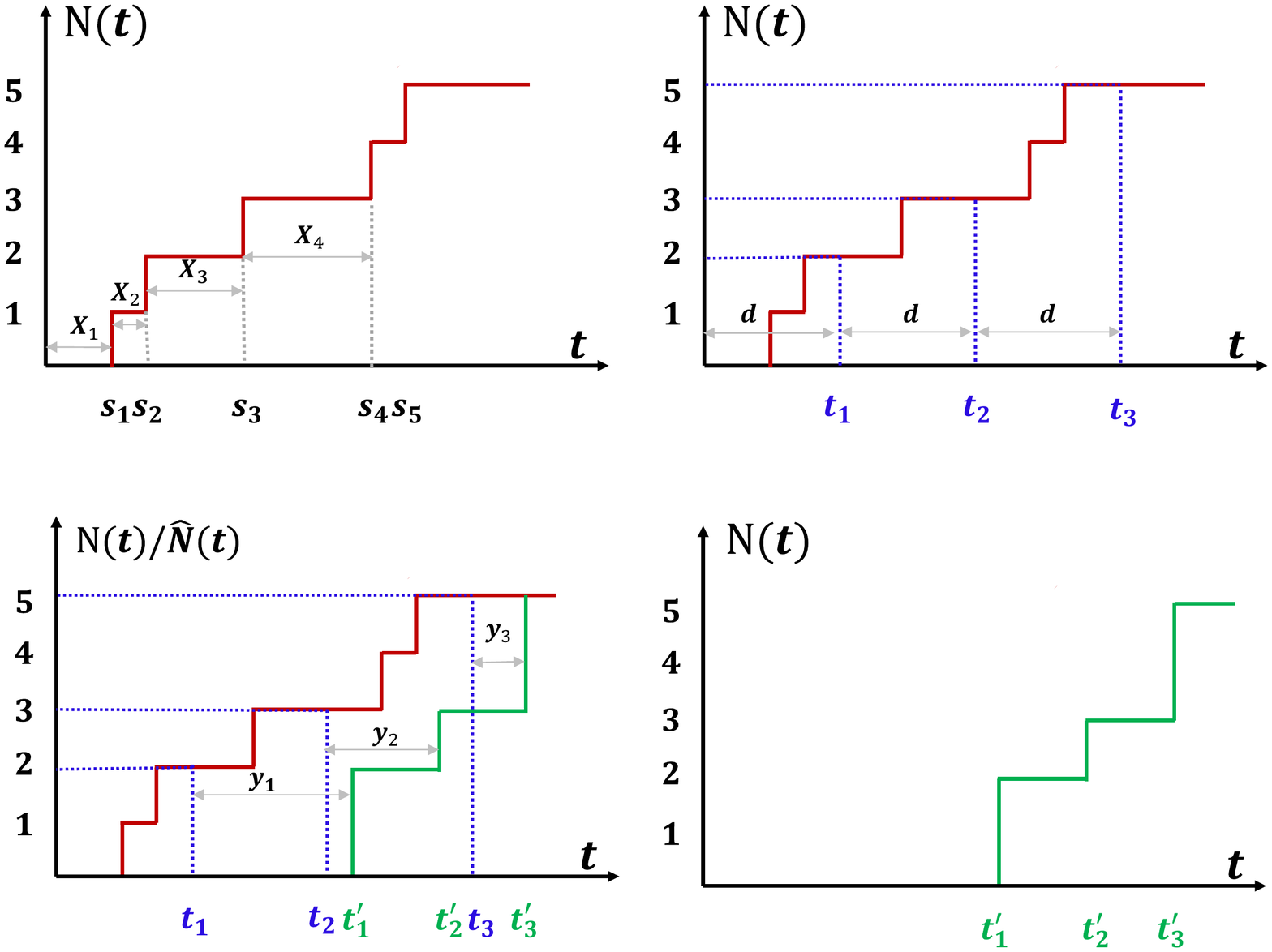}}
	\caption{The illustration of sampling and reconstruction}
	\label{1252311} 
	\vspace{-0.5cm}
\end{figure*}

In \figurename\ \ref{model:subfig:b}, we plot the first few events of process $N(t)$. The sampling is described in \figurename\ \ref{model:subfig:c}, where the uniform sampling is adopted as an example. In \figurename\ \ref{model:subfig:d}, we plot the reconstructed process as well as the original process. Combining these example pictures, we illustrate the system operation in details.

For counting process $N(t)$, let $\{s_n, n \geqslant 1\}$ denote the arrival time of the $n$th event and $s_0=0$. There is $s_n \leqslant s_{n+1}$, $\forall~ n \geqslant 0$. Let $\{X_n, n \geqslant 1\}$ be the sequence of interarrival times. We have $ X_n = s_{n} - s_{n-1}$, $\forall~ n \geqslant 1$. In this paper, we suppose that the original process $N(t)$ is a Poisson process with parameter $\lambda$. Thus, the interarrival time sequence $\{X_n, n \geqslant 1\}$ are independent identically distributed $(i.i.d.)$ exponential random variables with mean $\frac{1}{\lambda}$.

For the sampling process, we consider two different methods, i.e., the uniform sampling and the non-uniform sampling. Let $\{t_i, i \geqslant 1\}$ denote the sampling time sequence. For uniform sampling, the sampler works with $d$ $(d>0)$ being the sampling interval. Thus, the sampling time sequences are obtained as $\{t_i = id, i \geqslant 1\}$. For non-uniform sampling, the sampler is trigged by given events, such as a timer or a counter. For both methods, at time epoch $t_i$, the sampler measures process $N(t)$ and obtains the status of $N(t)$ at this time epoch, i.e., $N(t_i)$. Then, one packet is generated to denote the raw information of $N(t_i)$ and sent into a \mbox{server queue}\footnote{Suppose that the increase of $N(t)$ during one sampling interval can be exactly described by one packet. This assumption is ideal since the length of one packet is limited and the arrival of the original process during one sampling interval is possible infinite. More  detailed discussion can be found in \emph{Appendix} \ref{12011339}.}. 

For the server, we assume that the service time for each packet is $i.i.d.$ exponential random variable with mean $\frac{1}{\mu}$, which accounts for the wireless transmission time of each packet. The probability dense function $(p.d.f)$ of the service time is then given as
\vspace{-0mm}
\begin{align}\label{11121516}
f(x) = \mu e^{-\mu x}, \ x\geqslant 0.
\end{align}
\vspace{-0mm}
The packets in the queue wait for their turn to be served. After being served, these packets arrive at the monitor, where a new process $\hat{N}(t)$ is constructed to be an approximation of $N(t)$. Specifically, $\hat{N}(t)$ remains the value of last received sample and updates to a new value with a newly arrived sample, i.e.,
\vspace{-4mm}
\begin{align}
\hat{N}(t)=
\left\{
\begin{array}{ll}
N(t_{i-1}), & t < t_{i}^{'}; \\
N(t_{i}), & t \geqslant t_{i}^{'},
\end{array}
\right.
\vspace{-0mm}
\end{align}
where $t^{'}_i$ denote the arrival time of the $i$th sample. We assume the first-in-first-out queueing rule, i.e., $t^{'}_{i} < t^{'}_{i+1}$. As shown in \figurename\ \ref{model:subfig:d}, distortion occurs between the original process $N(t)$ and the reconstructed process $\hat{N}(t)$.
\vspace{-4mm}
\begin{definition}\label{def1}
Let $\hat{N}(t) = 0$, $t < t_1$. The real-time distortion $D(t)$ is defined as the difference between $N(t)$ and $\hat{N}(t)$, i.e.,
\vspace{-2mm}
\begin{align}
D(t) = N(t) - \hat{N}(t), ~t \geqslant 0.
\end{align}	
\end{definition}
\vspace{-2mm}	
Considering that $N(t)$ is a counting process and there exists non-negative delay between $N(t)$ and $\hat{N}(t)$, we conclude that $D(t) \geqslant 0$, $\forall~t \geqslant 0$. The real-time distortion occurs for two reasons. First, instead of recording the arrival time of every event of counting precess $N(t)$, the sampling omits the details of $N(t)$ during every sampling interval and simply records the result of the counting process at the sampling instant. Second, the sampled measurements cannot be received by the monitor without delay which occurs due to the waiting time and the service time. We consider the reconstruction during time horizon $[0,T]$.
 \vspace{-4mm} 
 \begin{definition}\label{def2}
	  The average distortion is defined as the $1$-$norm$ of function $D(t)$, namely, 
	 \begin{align}\label{5101505}
	 	\Theta = \frac{1}{T}\int_0^T D(t){\rm{d}}t.
	 \end{align}
 \end{definition}
\vspace{-5mm}

From \emph{Definition} \ref{def2}, the average distortion represents the average gap  between the two curves in \figurename\ \ref{model:subfig:d}. We evaluate the performance of the reconstruction with the average distortion in \eqref{5101505} being the criterion and try to minimize it by finding the optimal sampling parameters for  different sampling methods. 

\begin{table}[!t]
	\centering
	\caption{The average distortion under three sampling policies}
	\label{12011353}
	\begin{tabular}{|c|c|c|l|}
		
		\hline
		\multicolumn{2}{|c|}{Policy}&Parameter&Average distortion\\
		\hline
		\multicolumn{2}{|c|}{uniform sampling}& sampling rate $r$ & $\Theta(r) = \lambda\left(\frac{1}{2r} + \frac{\sigma}{\mu(1-\sigma)} + \frac{1}{\mu}\right)$\\
		\hline
		\multirow{2}{*}{non-uniform}& threshold-based policy & sampling threshold $\beta$ & $\Theta(\beta) = \lambda\left(\frac{\beta-1}{2\lambda} + \frac{1}{\mu(z_0^{\beta}-1)} + \frac{1}{\mu}  \right)$\\
		\cline{2-4}
		&	zero-wait policy & $\backslash$ &\  $\Theta_{zw}=\lambda \frac{2}{\mu}$ \\
		\hline
	\end{tabular}
\vspace{-5mm}	
\end{table}

As shown in \tablename \ \ref{12011353}, there are two kinds of method for the sampler to perform sampling. The first one, i.e., the uniform sampling, only observes the signal at some given time instants and reports the counting status at the corresponding time epochs. Another one corresponds to the non-uniform sampling, where the sampler keeps an eye on the counting process and generates a measurement to indicate the updates if given conditions are satisfied. This method can help the monitor know the original process timely but is costly since the sampler should always stay awake. The uniform sampling is much more economic since the sampler only needs to work at given time but it adds distortion in expressing the original process. In the following two sections, we consider the average distortion under the two sampling methods, respectively. Some notations in this paper are given in \tablename\ \ref{11301602} for convenience.

\begin{table}[!t]	
	\centering
	\caption{Notations in this paper}
	\label{11301602}
	\vspace{-3mm}
	\begin{tabular}{|c|c||c|c|}		
		\hline
		Symbol&Description &Symbol&Description\\
		\hline
		$N(t)$& the original process&$\hat{N}(t)$&the reconstructed process\\
		\hline
		$s_n$&the arrival time of $n$th event&$\{X_n\}$&the interarrival time sequence\\
		\hline
		$\lambda$&the arrival rate of $N(t)$&$\mu$& the service rate of queue server\\
		\hline
		$t_i$&the sampling time of $i$th sample&$t_i^{'}$&the receiving time of $i$th sample\\
		\hline
		$d$&the sampling interval&$r$& the  sampling rate\\
		\hline
		$\beta$&the sampling threshold&$f(x), p(x), h(x)$&$p.d.f.$s of random variables\\
		\hline
		$D(t)$&the reconstruction distortion&$\Theta, \breve{\Theta}, \Theta(\cdot), \Theta_{(\cdot)}$&the average distortion\\
		\hline
		$\sigma$&the solution of a Lambert W function&$z_0$&one of the zeroes of a polynomial\\
		\hline
		$S_{(\cdot)}$ or $\hat{S}_{(\cdot)}$&the area of a polygon&$\mathbb{E}\{\cdot\}$&the expectation operator\\
		\hline
	\end{tabular}
	\vspace{-3mm}
	
\end{table}

\vspace{-0mm}
\section{Uniform sampling method }
\vspace{-0mm}

In this section, we focus on minimizing the average reconstruction distortion when the uniform sampling method is adopted. In this case, the average distortion $\Theta$ is merely decided by the sampling rate $r $ which is defined as $r=\frac{1}{d}$ for a given counting process. Thus, we first formulate the closed-form expression of the average distortion-sampling rate function $\Theta(r)$, with which the optimal sampling rate can be obtained to minimize the average distortion. Then, we propose an algorithm that can further decrease the average distortion based on interpolation. At last, we give the lower bound of the average distortion when interpolation algorithms are adopted. 

\vspace{-0mm}
\subsection{The Average Distortion-Sampling Rate Function}\label{4241837}
\vspace{-0mm}

\begin{figure}
	\centering
	\includegraphics[width=0.7\columnwidth]{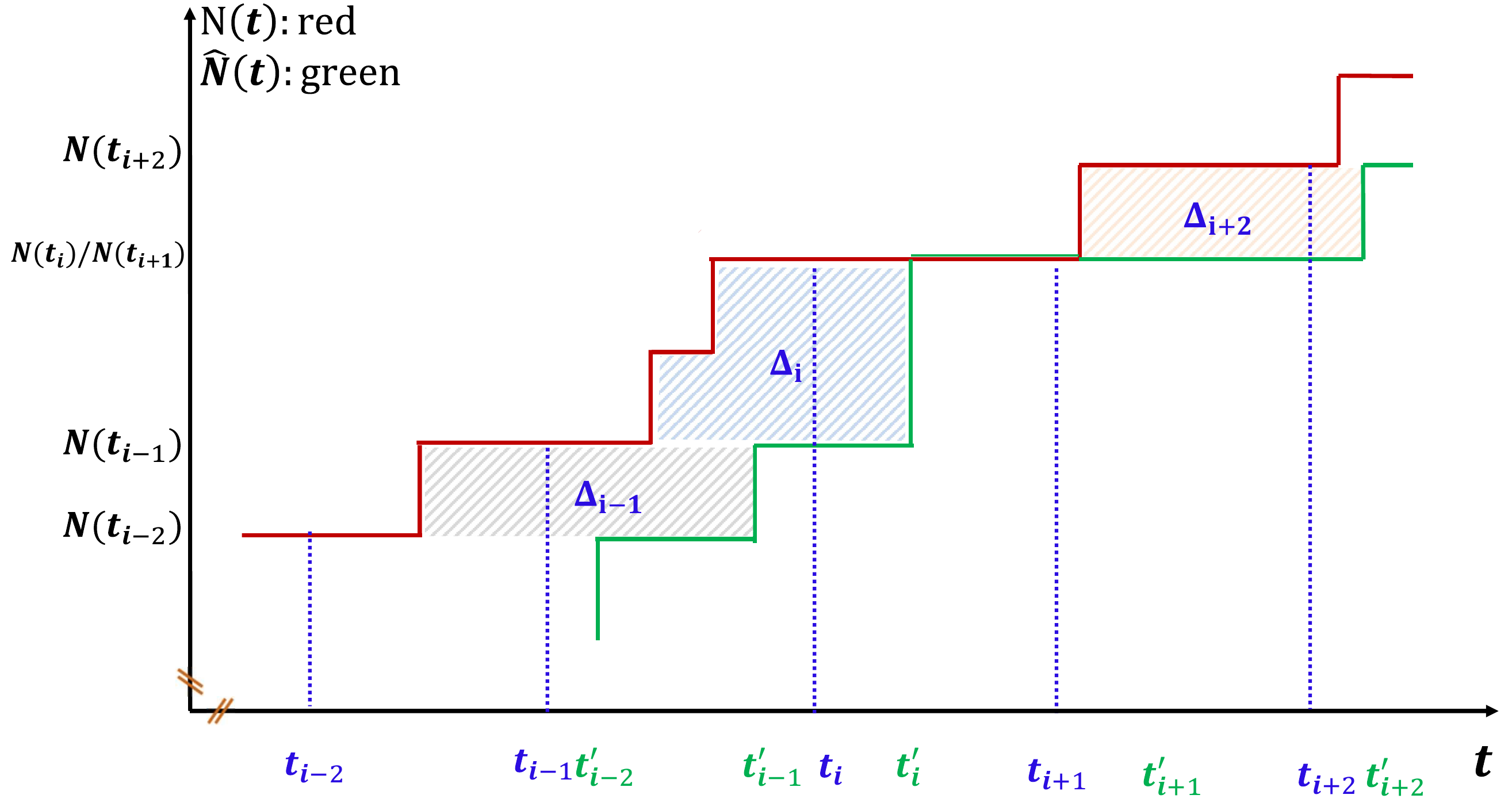}
	\setlength{\abovecaptionskip}{-0mm}
	\caption{Distortion under the  uniform sampling policy}
	\label{1271428}
	\vspace{-0mm}
\end{figure}

As shown in \figurename\ \ref{1271428}, the overall distortion area can be divided into $I = \frac{T}{d}$ polygons\footnote{Suppose that $T$ can be exactly divided by $d$. } which are marked as $\{\Delta_i, 1\leqslant i \leqslant I\}$ as the shaded areas. In detail, polygon $\Delta_i$ is circled by curve  $\{N(t), t_{i-1}\leqslant t \leqslant t_i\}$, line that starts from point $\left(t_{i-1}, N(t_{i-1})\right)$ and ends at point $(t_i^{'}, N(t_{i-1}))$, line that starts from point $(t_i^{'}, N(t_{i-1}))$ and ends at point $(t_i^{'}, N(t_{i}))$, and line that starts from point $(t_i, N(t_{i}))$ and ends at point $(t_i^{'}, N(t_{i}))$. Notice that, the area of $\Delta_i$ can be zero if no event happen during sample interval $(t_{i-1},t_{i}]$, such as $\Delta_{i+1}$ in \figurename\ \ref{1271428}.

\vspace{-0mm}
\begin{lemma}\label{lemma1}
	The average distortion-sampling rate function under uniform sampling is given by 
	\vspace{-0mm}
	\begin{align}\label{12011429}
	\Theta(r) = r\mathbb{E}\{S_{\Delta}\},
	\end{align}
	\vspace{-2mm}
	where $S_{\Delta_i}$ is the area of $\Delta_i$ and $\mathbb{E}\{S_{\Delta}\}$ is the average of $S_{\Delta_i}$.
\end{lemma}
\begin{IEEEproof}
	Based on the definition in \eqref{5101505}, the average distortion is the average gap between the two curves. Hence with the division in \figurename\ \ref{1271428}, it can be formulated as 
	\begin{align}
	\Theta(r) &= \frac{1}{T}\int_0^T D(t){\rm{d}}t
	= \frac{I}{T}\frac{1}{I}\sum\limits_{i = 1}^{I}S_{\Delta_i}.
	\label{211602}
	\end{align}
	In \eqref{211602}, the item $\frac{I}{T}$ is the sampling rate $r$ and the item $\frac{1}{I}\sum\nolimits_{i = 1}^{I}S_{\Delta_i} $ is the average area of the divided polygons. Thus, we obtain the average distortion in \eqref{12011429}. 
	\end{IEEEproof}

\vspace{-0mm}	
\begin{lemma}\label{lemma2}
	The average area of the divided polygons is given as
	\begin{align}
	\mathbb{E}\left\{S_{\Delta}\right\}  =  \frac{\lambda}{2r^2} + \frac{\lambda}{r} \left( \frac{\sigma}{\mu(1-\sigma)} + \frac{1}{\mu} \right),
	\label{3131106}
	\end{align}
	where parameter $\sigma$ is the solution of Lambert W Function, namely,
	\begin{align}\label{12011602}
	\sigma = -\frac{r}{\mu}\mathcal{W}(-\frac{\mu}{r}e^{-\frac{\mu}{r}}).
	\end{align} 
\end{lemma}	
\vspace{-0mm}

\begin{figure}[t]
	\centering
	\includegraphics[width=0.7\columnwidth]{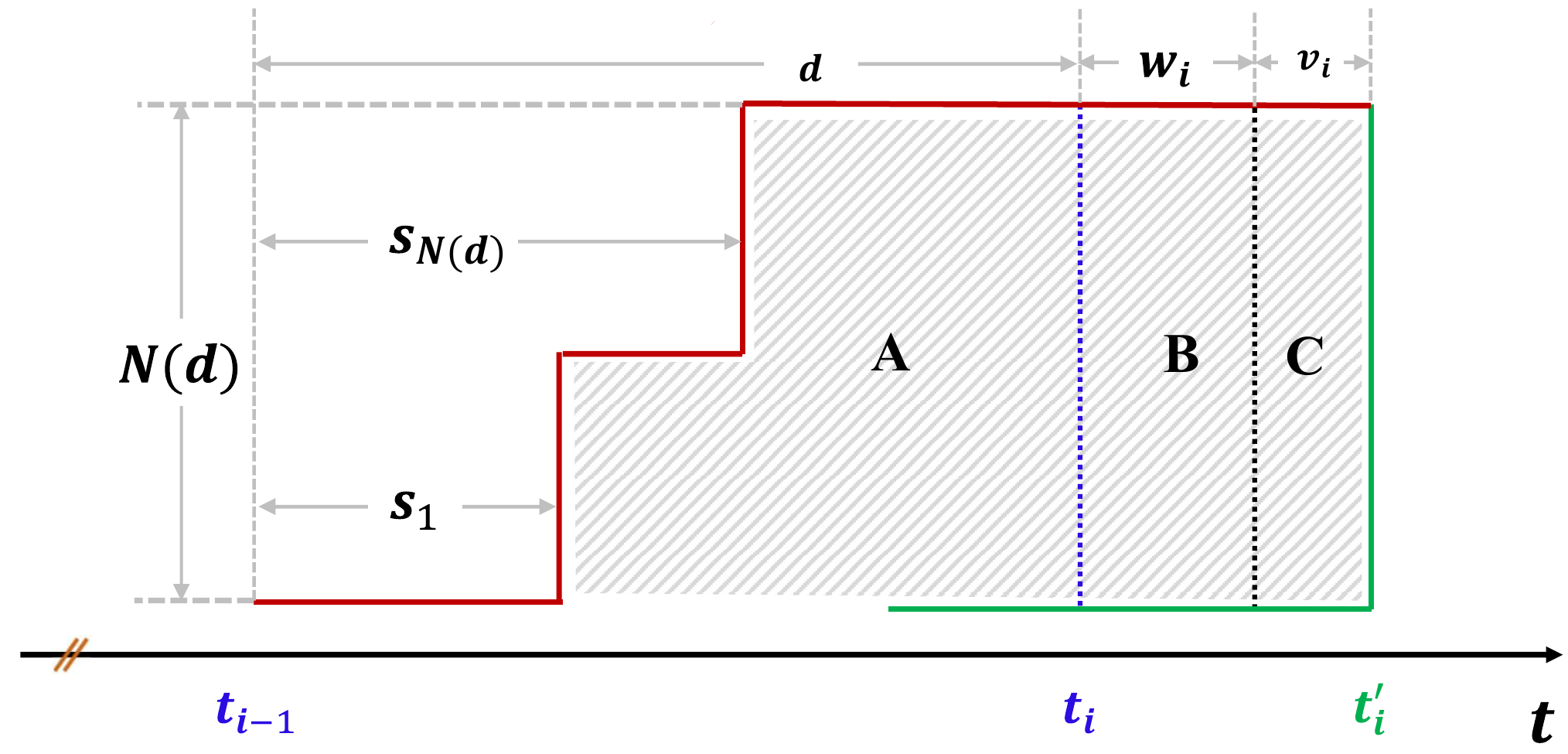}
	\vspace{-0.0cm}
	\setlength{\abovecaptionskip}{-0mm}
	\caption{An example of $\Delta_i$: $d$ is the average sampling interval, $N(d)$ is the height of the polygon (the number of events occurring in the sampling interval), $w_i$ is the waiting time of sample $i$ in the queue and $v_i$ is the service time of the $i$th sample.}
	\label{1271639}
	\vspace{-0.0cm}
\end{figure}

\begin{IEEEproof}
As shown in \figurename\ \ref{1271639}, by analyzing the structure of the polygon, we can divide it into three sub-polygons marked as $A$, $B$, and $C$ based on time (horizontal axis) that correspond to the distortions caused by sampling, waiting time, and service time. Specifically, starting from $t_{i-1}$ and ending at $t_{i}$, sub-polygon $A$ is the distortion caused by sampling; starting from $t_i$ and ending at $(t_i+w_i)$, sub-polygon $B$ is the distortion caused by queueing delay; and starting at $(t_i+w_i)$ and ending at $(t_i+w_i+v_i)$, sub-polygon $C$ is the distortion caused by the service time. Thus, we have
\vspace{-0mm}
\begin{align}\label{12011545}
\mathbb{E}\{S_{\Delta}\}  &= \mathbb{E}\{S_A\} + \mathbb{E}\{S_B\} + \mathbb{E}\{S_C\},
\vspace{-0mm}
\end{align}
where $S_A$, $S_B$, and $S_C$ represents the area of sub-polygons $A$, $B$, and $C$, respectively. It means we are able to calculate $\mathbb{E}\{S_{\Delta}\}$ as the sum of the average areas of the three divided sub-polygons, which are given in \emph{Appendix} \ref{12011542} for brevity.
\end{IEEEproof}

\vspace{-0mm}
\begin{theorem}\label{theorem1}
	The average distortion-sampling rate function is given as
	\begin{align}\label{eq1}
	\Theta(r) =  \lambda \Big( \frac{1}{2r} +  \frac{\sigma}{\mu(1-\sigma)} + \frac{1}{\mu}  \Big). 
	\end{align}
\end{theorem}
\vspace{-0mm}
\begin{IEEEproof}
	Combining the results in \emph{Lemma} \ref{lemma1} and \emph{Lemma} \ref{lemma2}, we obtain the expression of the average distortion-sampling rate function in \eqref{eq1}.
\end{IEEEproof}

In \emph{Theorem} \ref{theorem1}, we give the average distortion-sampling rate function. The optimal sampling rate is the one that gives the minimum average distortion, i.e.,
\vspace{-0mm}
\begin{align}\label{4231134}
r^* = \mathop{\arg\min}_{r} \quad \Theta(r).
\end{align}
\vspace{-0mm}
Since $\Theta(r)$ only has one variable, i.e., the sampling rate $r$, one can find the optimal sampling rate $r^{*}$ by calculating its derivation and making it as zero. However, it is not explicit to obtain the derivation due to the existing of parameter $\sigma$. In the simulation section, we can plot the curve of function $\Theta(r)$ and choose a sampling rate that induces a smaller average distortion. 

In the end, we consider the optimal sampling rate from another point of view. As we have illustrated that the distortion occurs due to sampling and transmission. These two kinds of distortion occur simultaneously and twist with each other\footnote{One can imagine that, for the reconstructed process $\hat{N}(t)$, it only suffers from shape change if the reconstruction is done locally or only suffers from time translation if the sampler always keeps an eye on $N(t)$ and updates for every count.}. Thus,  there exists an interesting tradeoff between these two kinds of distortion  when we adjust the sampling rate. Specifically, if we sample with a smaller $r$, the distortion caused by sampling will increase since the description of $N(t)$ is less elaborate, while the distortion caused by transmission will decrease since less packets will arrive at the queue. On the contrary, if we sample with a greater $r$, the distortion cased by sampling will decrease while the distortion caused by transmission will increase since more arrival packets lead to higher waiting time. The optimal sampling rate $r^*$ is the rate than balances this tradeoff.
\vspace{-0mm}
\subsection{Reconstruction with Interpolation Algorithms}
\vspace{-0mm}

\begin{figure}[t]
	\centering
	\includegraphics[width=0.7\columnwidth]{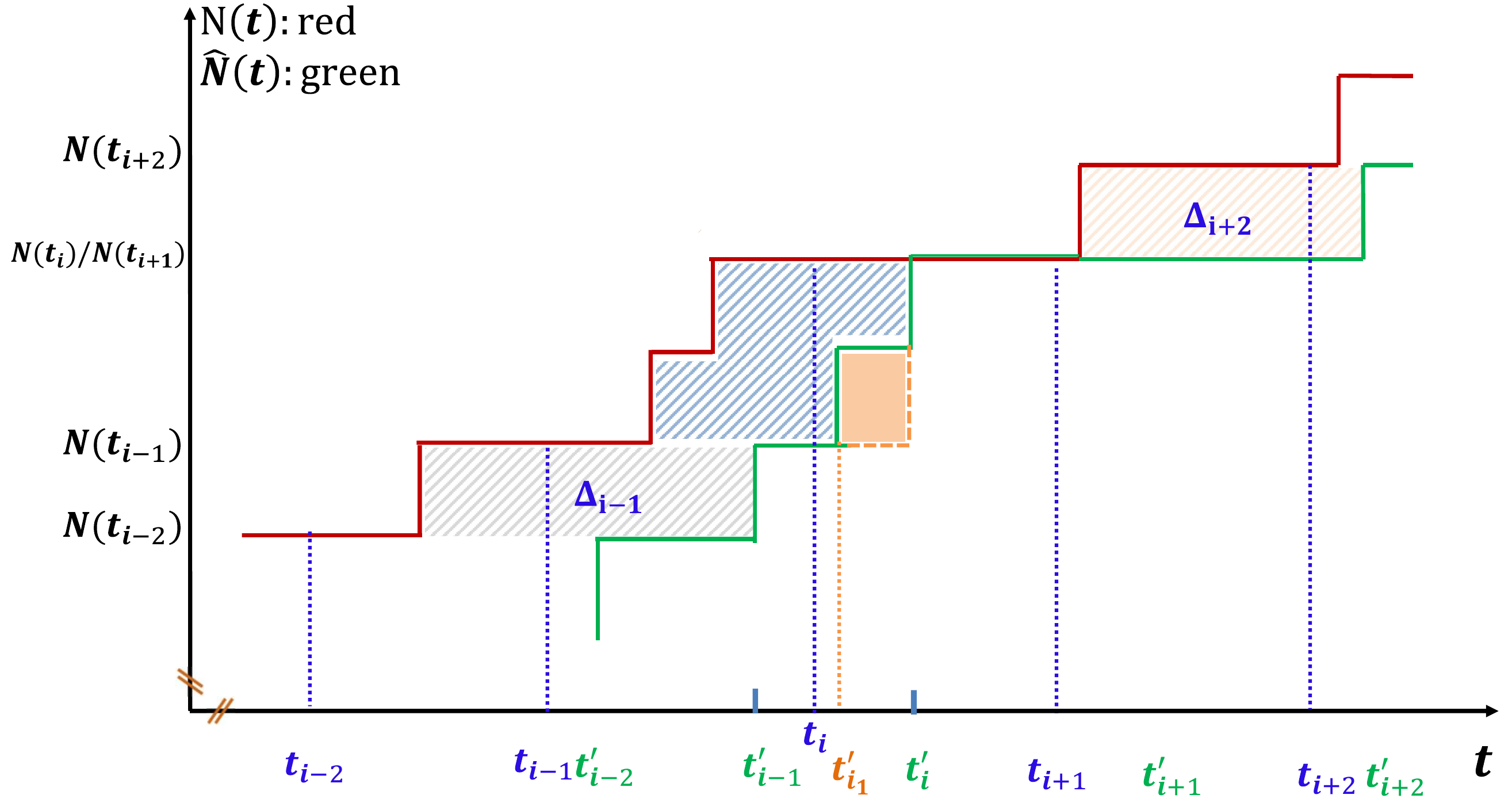}
	\vspace{-0mm}
	\setlength{\abovecaptionskip}{-0mm}
	\caption{Reconstruction with guessing: the orange area is the decreased distortion by adding one guessing point.}
	\label{1291008}
	\vspace{-0mm}
\end{figure}

In this subsection, we first propose interpolation algorithms to further decrease the average distortion. Then, the lower bound of the average distortion is provided as the upper bound of the performance of these interpolation algorithms. 

\subsubsection{Interpolation Algorithms} These algorithms are based on the following idea: although the sampling omits the details of the original process during the sampling interval, the monitor still can make some estimations for the omitted details, i.e., reconstruction with interpolation.

By interpolation, we mean that, if the monitor receives one packet at time $t_i^{'}$ which indicates any increase compared to the last received packet at time $t_{i-1}^{'}$, the monitor can insert some updates itself during interval $(t_{i-1}^{'}, t_i^{'})$ to decrease the reconstruction  distortion. An example of this idea is given in \figurename\ \ref{1291008}, where the new received packet at time $t_{i}^{'}$ indicates that the original process is increased by $2$ compared to the last sample, but we don't know exactly when do these events occur. In this case, estimation can be made for the occurring time. For example, at time $t_{i_1}^{'}$, the reconstructed process $\hat{N}({t_{i_1}^{'}})$ is set to $N(t_{i-1}^{'}) + 1$. As a result, the overall distortion is decreased by the area labeled with orange color. In \emph{Algorithm} \ref{4231925}, we give an algorithm that adopts uniform guessing for the omitted details, which means that the inserted points are chosen to be uniformly distributed on the open interval $(t_{i-1}^{'}, t_i^{'})$.

Except for \emph{Algorithm} 1, there are many algorithms that can be carefully designed  to further decrease the average distortion. However, these algorithms only decrease the sampling distortion since we are trying to recover the details omitted due to sampling. Thus, the total distortion cannot be eliminated.

\begin{spacing}{1.1}
	\begin{algorithm}[t]   
		\caption{ Reconstruct $\hat{N}(t)$ with interpolation}   
		\label{4231925}    
		\begin{algorithmic}[1] 
			\REQUIRE ~~\\   
			the time horizon $T$;  
			\ENSURE ~~\\  
			the reconstructed process $\hat{N}(t)$; 
			\STATE{Initialize: $\hat{N}(0) = 0$};
			\WHILE{receives a packet at time $t_i^{'}$ and $t_i^{'} < T$}  
			\STATE Sets $\hat{N}(t_i^{'}) = N(t_i)$ and calculates $ J = \hat{N}(t_i^{'}) - \hat{N}(t_{i-1}^{'}) - 1$;
			\IF {$ J > 0$} 
			\STATE uniformly generates $J$ real numbers $\{t_{i_j}^{'} | 1\leqslant j\leqslant J \}$ from open interval $(t_{i-1}^{'}, t_{i}^{'})$;
			\FOR {every $j$ in set $\{1,2,\cdots, J\}$ }
			\STATE  $\hat{N}(t_{i_1}^{'})$ = $\hat{N}(t_{i-1}^{'})$ + $j$;
			\ENDFOR
			\ENDIF 
			\STATE  $i = i + 1$;
			\ENDWHILE 
		\end{algorithmic}  
	\end{algorithm}  
\end{spacing}

\subsubsection{The Lower Bound of the Average Distortion with Interpolation} We then give the upper bound of the performance of these interpolation algorithms, namely, the lower bound of the average distortion.

\vspace{-0mm}
\begin{theorem}
	The lower bound of the average distortion under interpolation algorithms is \mbox{given as}
	\begin{align}	\label{12011708}
	\breve{\Theta} = r\sigma\mu(1-\sigma){\rm{e}}^{-\frac{\lambda}{r}}
	\sum\limits_{n=1}^{\infty}\frac{\lambda^n}{n!r^n} &\left(n\int_{\frac{1}{r}}^{\infty}(x-\frac{1}{2r}){\rm{e}}^{-\mu(1-\sigma)x}{\rm{d}}x  \right. \nonumber\\
	& \left. + \frac{1}{2}\int_{0}^{\frac{1}{r}} x {\rm{e}}^{-(\lambda+\mu-\mu\sigma)x} \sum\limits_{m=1}^{\infty} \frac{(\lambda x)^m}{(m-1)!}{\rm{d}}x\right).
	\end{align}
\end{theorem}
\begin{IEEEproof}
The principle that achieves	$\breve{\Theta}$ is that, making the curve after interpolation approach the original process as much as possible such that the distortion area is decreased to the most extent. As shown in \figurename\ \ref{10102206}, this principle requires different operations for different cases.

\begin{figure}[t]
	\centering
	\subfigure[\emph{Case} 1: $t_{i-1}^{'} \geqslant t_i$]{
		\vspace{-0mm}
		\label{sketch1} 
		\includegraphics[width=0.45\columnwidth]{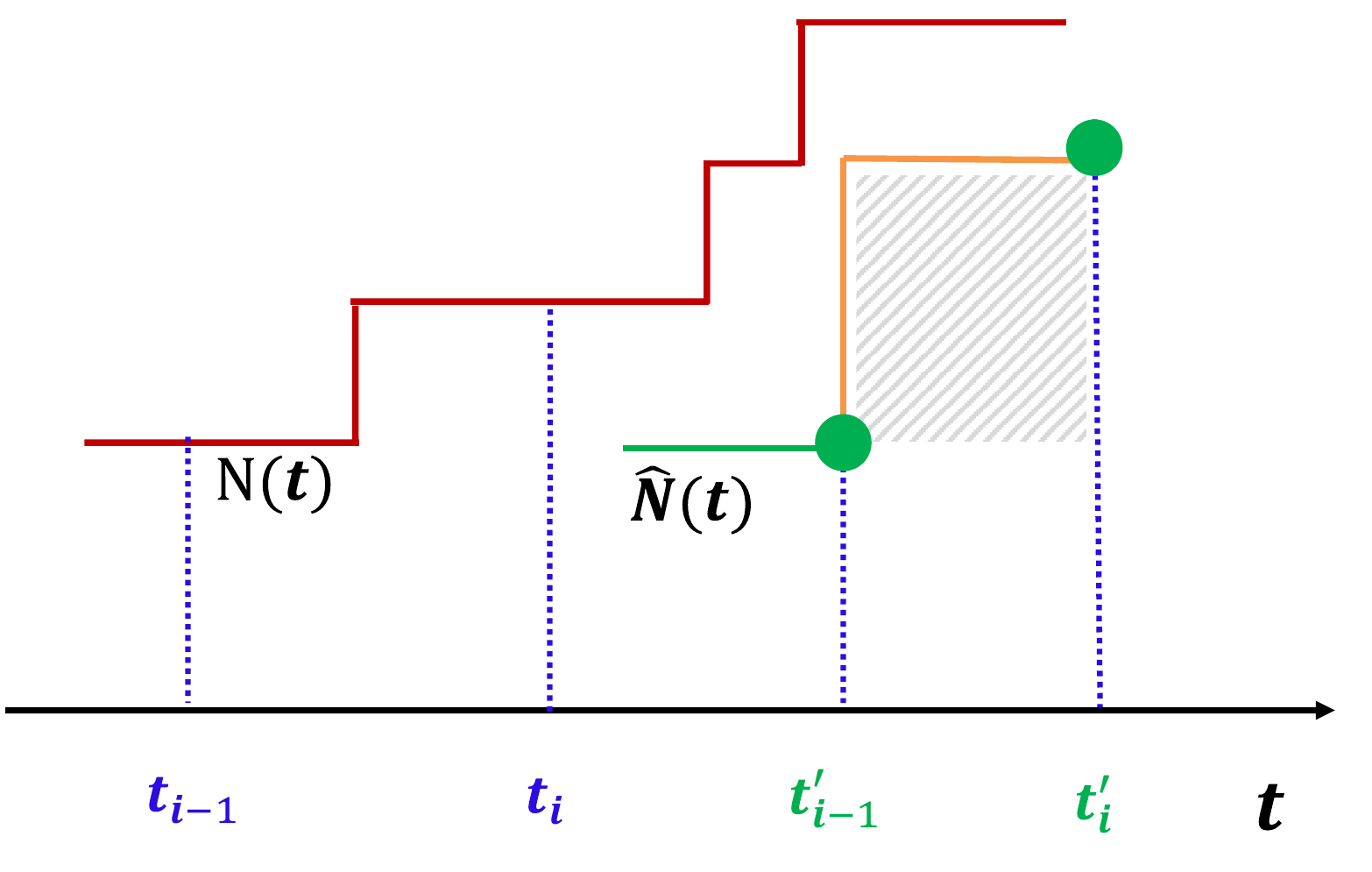}
		\vspace{-0mm}
	}
	\subfigure[\emph{Case} 2: $t_{i-1}^{'} < t_i$]{
		\vspace{-0mm}
		\label{sketch2} 
		\includegraphics[width=0.45\columnwidth]{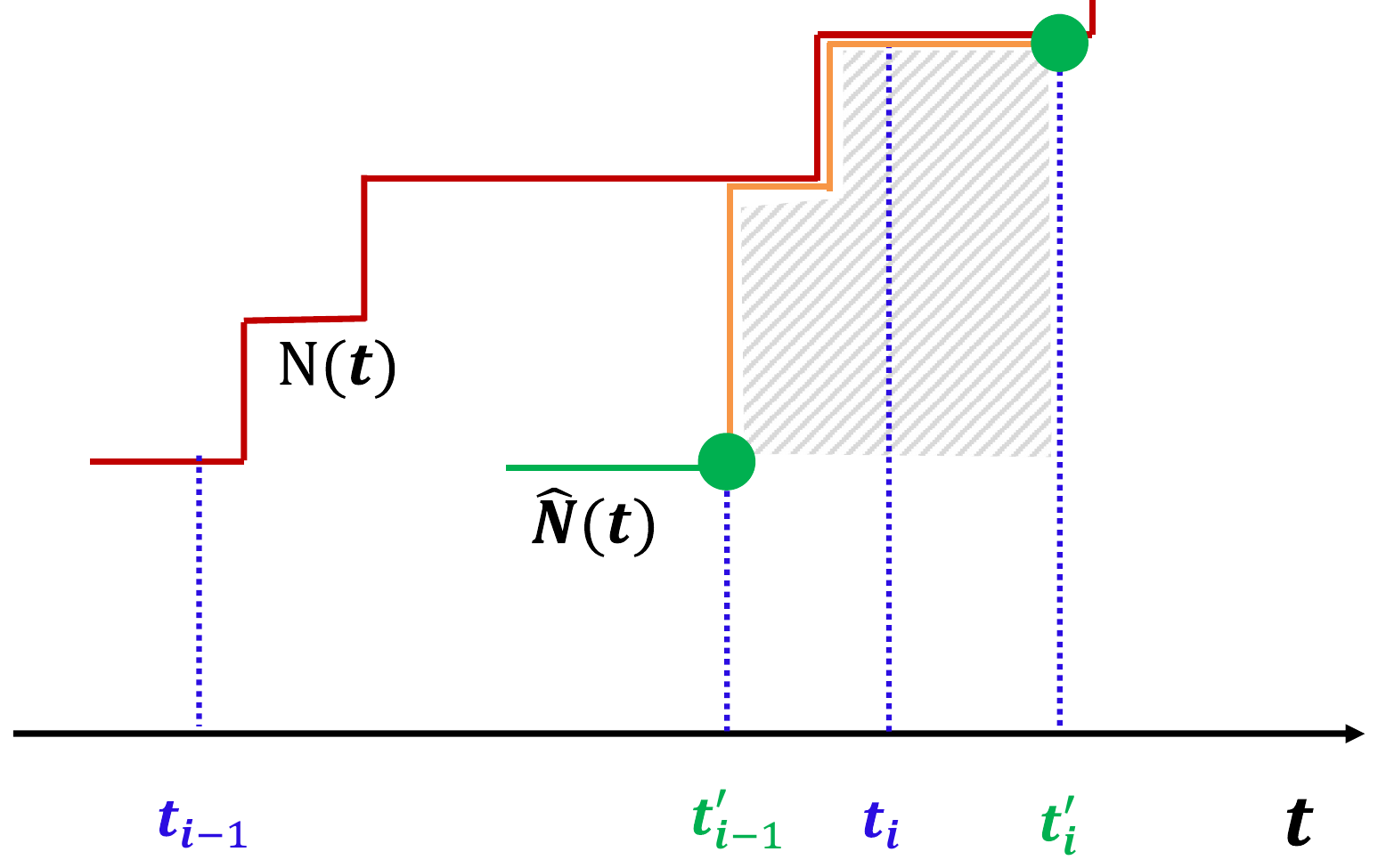}
		\vspace{-0mm}
	}	
	\vspace{-0mm}
	\caption{Sketches for \emph{Case} 1 and \emph{Case} 2}
	\label{10102206} 
	\vspace{-2mm}
\end{figure}

\emph{Case 1:} The received time of the $(i\!-\!1)$th sample $t_{i-1}^{'}$ and the sampling time of the $i$th sample $t_i$ meet the requirement  $t_{i-1}^{'} \geqslant t_i$. The minimal average area of one polygon in this case is calculated as\footnote{The derivations for conclusions in \eqref{10291505} and \eqref{10291504} are given in \emph{Appendix} \ref{12011714} due to space limitation.}
\vspace{-0mm}
\begin{align}\label{10291505}
\hat{S}_1 = \sum\limits_{n=1}^{\infty} n\text{Pr}\{N(d)=n\} \int_d^\infty p(x) (x-\frac{d}{2}) {\rm{d}}x,
\end{align}
where $p(x)$ is the $p.d.f$ of the random variable defined as the sum of the waiting time and the service time, and given as \cite{kleinrock1975queueing}
\vspace{-0mm}
\begin{align}\label{12011707}
p(x) = \sigma\mu(1-\sigma){\rm{e}}^{-\mu(1-\sigma)x}, \ x \geqslant 0.
\vspace{-0mm}
\end{align}
In \emph{case} $1$, the best reconstruction occurs if we increase the constructed process \mbox{as soon as possible.} 

\emph{Case 2:} The received time of the $(i\!-\!1)$th sample $t_{i-1}^{'}$ and the sampling time of the $i$th sample $t_i$ meet the requirement $t_{i-1}^{'} < t_i$. The minimal average area of one polygon in this case is calculated as
\vspace{-2mm}
\begin{align}\label{10291504}
\hat{S}_2 = \sum\limits_{n=1}^{\infty}\!\text{Pr}\{N(d)=n\} \int_0^d \frac{x}{2}\sum\limits_{m=1}^{n}m\text{Pr}\{N(x)=m\}  p(x) {\rm{d}}x.
\vspace{-2mm}
\end{align}
In \emph{case} $2$, the best reconstruction occurs if we catch up with the original process as soon as possible and then keep pace with it.

Combining the two cases, we can obtain $\breve{\Theta} = r(\hat{S}_1+\hat{S}_2)$ as
\begin{align}
\breve{\Theta} \!=\! r\sum_{n=1}^{\infty} \text{Pr}\{N(d)\!=\!n\}\left(n\int_{d}^{\infty}p(x)(x-\frac{d}{2}){\rm{d}}x + \frac{1}{2}\int_{0}^{d} xp(x)\sum_{m=1}^{n}m\text{Pr}\{N(x)=m\} {\rm{d}}x \right)
\end{align}
as the lower bound for average distortion when interpolation is adopted during reconstruction. Recall that, $\text{Pr}\{N(d)=n\} = \frac{(\lambda d)^n}{n!}{\rm{e}}^{-\lambda d}$,  $\text{Pr}\{N(x)=m\}  = \frac{(\lambda x)^m}{m!}{\rm{e}}^{-\lambda x}$, and the $p(x)$ in \eqref{12011707}, we arrive at the conclusion in \eqref{12011708}.
\end{IEEEproof}

The specific strategy that achieves lower bound $\breve{\Theta}$ is ideal and impractical, however, it can be used to evaluate any given interpolation algorithm.
%
\vspace{-0mm}
\section{Non-uniform sampling method}
\vspace{-0mm}
In this section, we focus on minimizing the average reconstruction distortion when non-uniform sampling methods are adopted. Especially, we give the closed-form expressions for the average distortion under the threshold-based policy and the zero-wait policy.

For the non-uniform sampling, we have to tell the sampler exactly at what time epochs it performs sampling. Thus, contrasting to the uniforming sampling, the sampler has to keep awake and observes the "trigger". Once the trigger event occurs, it immediately performs one sampling. Considering the maneuverability, we focus on two sampling policies, namely, the threshold-based sampling and the zero-wait sampling. By threshold-based sampling, we mean that the sampler only obtains one sample when the original process counting is accumulated to threshold value $\beta$\footnote{The uniform sampling can be regarded as a threshold-based sampling policy with threshold imposed on the time dimension.}. The trigger event is "there are $\beta$ events occurring". By zero-wait sampling, we mean that the sampler obtains one sample only when the server is idle. The trigger event is "the server is idle". In the following, we study the average distortions under the threshold-based policy and zero-wait policy in sequence.
\vspace{-1mm}
\subsection{Threshold-Based Sampling policy}
\begin{figure}[t]
	\centering
	\subfigure[System model for the  threshold-based sampling policy]{
		\vspace{-0mm}
		\label{5101109_a} 
		\includegraphics[width=0.95\columnwidth]{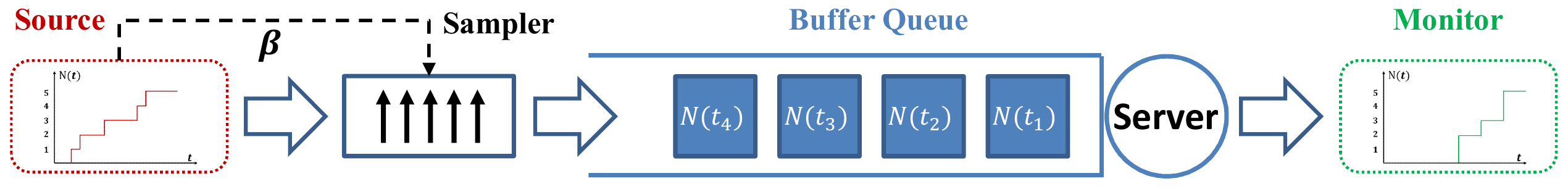}
		\vspace{-0mm}
	}\\
	\subfigure[The average distortion]{
		\vspace{-0mm}
		\label{5101109_b} 
		\includegraphics[width=0.45\columnwidth]{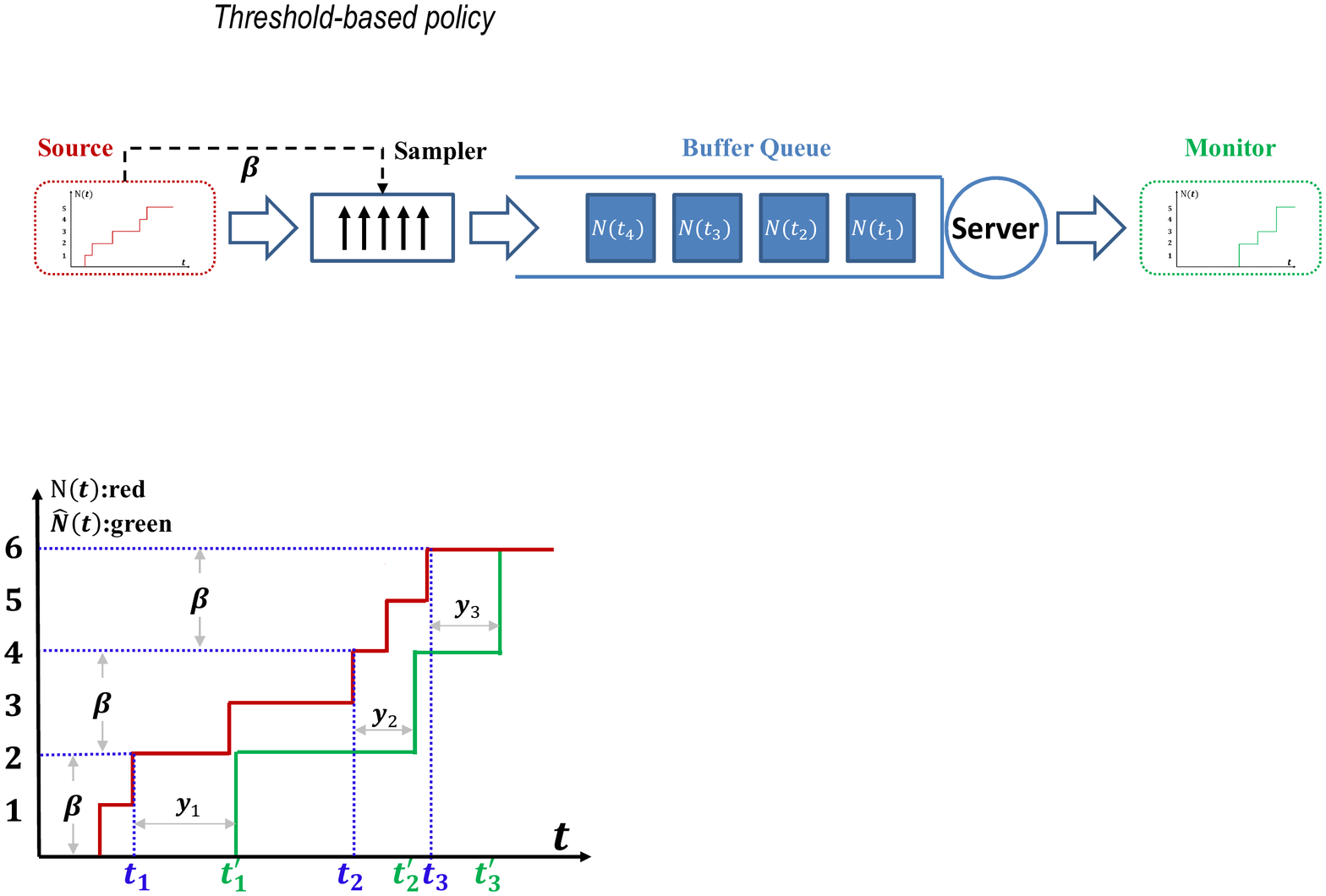}
		\vspace{-0mm}
	}	
	\subfigure[An example of the divided polygon]{
		\vspace{-0mm}
		\label{5101109_c} 
		\includegraphics[width=0.48\columnwidth]{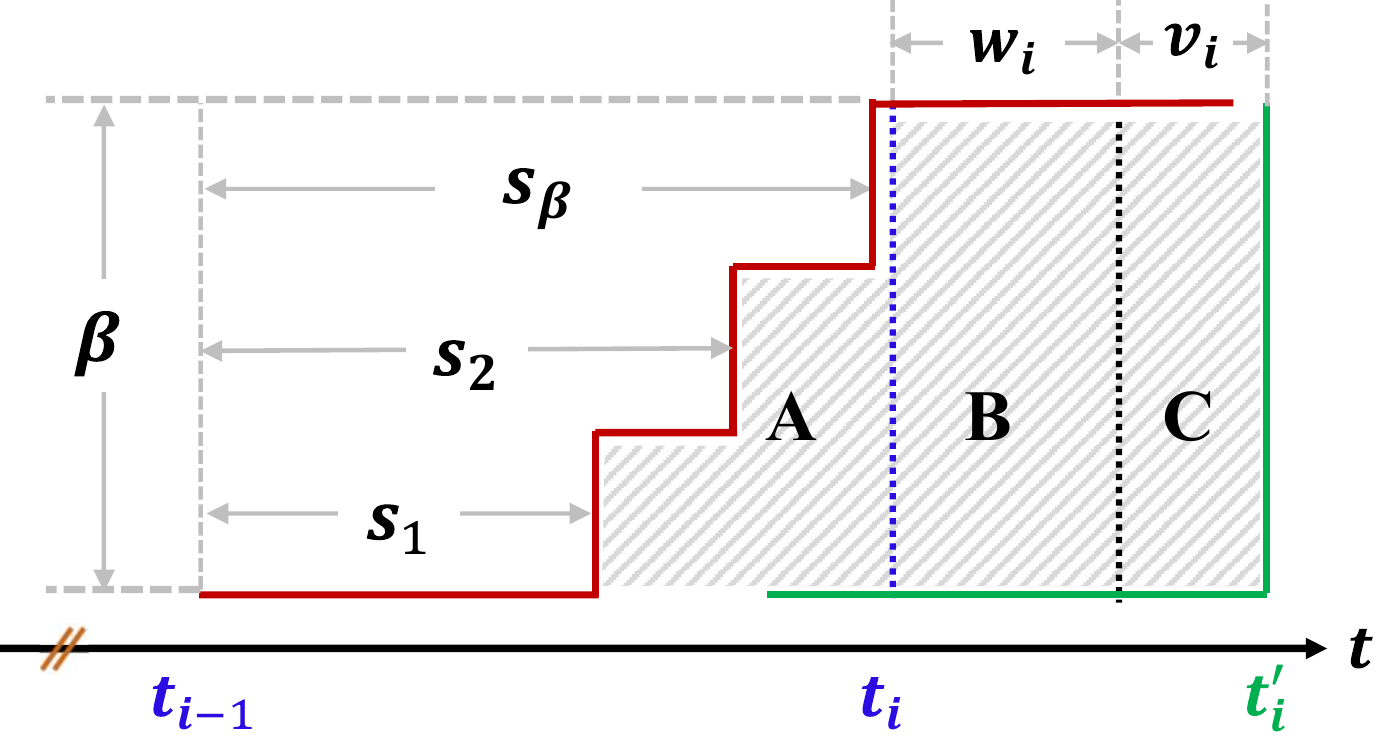}
		\vspace{-0mm}
	}	
	\caption{The system model, the distortion curve, and the divided polygons under threshold-based policy}
	\label{5101109} 
	\vspace{-0.6cm}
\end{figure}

\vspace{-1mm}
By threshold-based sampling, the sampler only samples when the accumulation of the original process equals or exceeds the given threshold $\beta$ ($\beta \geqslant 0$), as shown in \figurename\ \ref{5101109_a}. The threshold $\beta$ is a key parameter of the threshold-based sampling, since it determines the frequency of sampling and the traffic load of the queueing system, which are two critical factors of the reconstruction distortion. In this subsection, we focus on deriving the closed-form expression of the average distortion-sampling threshold function.

In \figurename\ \ref{5101109_b}, we give the example curves for the original and the reconstructed processes. The gap between $N(t)$ and $\hat{N}(t)$ is the overall distortion. It can be divided into $I$ polygons, where $I$ is the total sampling numbers. Especially, the $i$th polygon is denoted by $\Delta_i$ and circled by curve $\big\{N(t), t_{i-1} \leqslant t \leqslant t_i\big\}$, line $\Big[\big(t_i, N(t_i)\big), \big(t_i{'}, N(t_i)\big)\Big]$, line $\Big[\big(t_i^{'}, N(t_{i-1})\big), \big(t_i^{'}, N(t_{i}) \big)\Big]$, and line $\Big[\big(t_{i-1}, N(t_{i-1}), \big(t_i^{'}, N(t_{i-1}) \big)\Big]$ as shown in \figurename\ \ref{5101109_c}. 

\begin{lemma}\label{lemma3}
	The average distortion under the threshold-based sampling policy is reformulated as
	\begin{align}
	\Theta(\beta) = \frac{\lambda}{\beta}\mathbb{E}\left\{S_{\Delta}\right\},\label{5101518}
	\end{align}
	where $S_{\Delta_i}$ is the area of $\Delta_i$ and $\mathbb{E}\left\{S_{\Delta}\right\}$ is the average of $S_{\Delta_i}$.
\end{lemma}
\begin{IEEEproof}
Based on \eqref{5101505}, the average distortion $\Theta(\beta)$ can be rewritten as
\begin{align}
\Theta(\beta)  =  \frac{1}{T}\int_0^T D(t){\rm{d}}t  = \frac{I}{T}\frac{1}{I}\sum\limits_{i = 1}^{I} S_{\Delta_i}.\label{5101510}
\end{align}
In \eqref{5101510}, the term $\frac{I}{T}$ can be rewritten as $\frac{\lambda}{\beta}$ since samples are obtained for every $\beta$ events and the term $\frac{1}{I}\sum\nolimits_{i = 1}^{I} S_{\Delta_i}$ is the average area of the divided polygons.
\end{IEEEproof} 
\vspace{-2mm}
\begin{lemma}\label{lemma4}
	The average area of the polygons is given as
	\begin{align}\label{5311051}
	\mathbb{E}\left\{S_{\Delta}\right\} = \beta \left( \frac{\beta-1}{2\lambda} + \frac{1}{\mu(z_0^{\beta}-1)} + \frac{1}{\mu}\right),
	\end{align}
	where $z_0$ is one of the zeroes of polynomial $\frac{\lambda}{\mu} z^{\beta+1} - (1+\frac{\lambda}{\mu}) z^{\beta} + 1$, whose modulus is greater than one, i.e., $|z_0| > 1$\footnote{There is only one zero whose modulus is greater than one, namely, $z_0$, there is one zero whose modulus is equal one, and modulus of the left $\beta-1$ zeroes are less than one.}.
\end{lemma}
\vspace{-4mm}
\begin{IEEEproof}
	As shown in \figurename\ \ref{5101109_c}, a polygon can be divided into three sub-polygons denoted by $A$, $B$, and $C$, that account for the sampling distortion, the distortion caused by sampling, and the distortion caused by service time. Thus, the average of $S_{\Delta}$ can be calculated as $ \mathbb{E}\left\{S_{\Delta}\right\} = \mathbb{E}\left\{S_A\right\} + \mathbb{E}\left\{S_B\right\} + \mathbb{E}\left\{S_C\right\}$. Specifically, we calculated $\mathbb{E}\left\{S_A\right\}=\frac{(\beta-1)\beta}{2\lambda}$ and $\mathbb{E}\left\{S_B\right\}=\frac{\beta}{\mu(z_0^{\beta}-1)}$, whose derivations are given in \emph{Appendix} \ref{12012359} for brevity. And we have $\mathbb{E}\left\{S_C\right\}=\beta \mathbb{E}\left\{v_i\right\} = \frac{\beta}{\mu}$. Summing these three items up, we arrive at the conclusion in \eqref{5311051}.
	\end{IEEEproof}
\vspace{-0mm}
\begin{theorem}
	The average distortion-sampling threshod function is give as
	\begin{align}\label{10281146}
	\Theta(\beta) = \frac{\lambda}{\beta}\mathbb{E}\left\{S_{\Delta}\right\}
	= \lambda\left(\frac{\beta-1}{2\lambda} + \frac{1}{\mu(z_0^{\beta}-1)} + \frac{1}{\mu}\right).
	\end{align}
\end{theorem}
\vspace{-0mm}
\begin{IEEEproof}
	Combining the results in \eqref{5101518} and \eqref{5311051}, we obtain the conclusion.
\end{IEEEproof}

The optimal threshold is the the one that leads to the minimum average distortion, i.e.,
\vspace{-0mm}
\begin{align}
\beta^*=\mathop{\arg\min}_{\beta}\  \Theta(\beta).
\end{align}
\vspace{-4mm}

Similar to the relationship between the average distortion and the sampling rate for uniform sampling, there exists similar relationship between the average distortion and the sampling threshold. That is, a greater threshold will evidently decrease the traffic load of the queueing system, as well as the queueing delay experienced by the samples. Thus, the distortion caused by transmission will be decreased. At the mean time, this greater threshold will increase the distortion caused by sampling. There is similar result for a smaller threshold. Thus, the optimal sampling threshold $\beta^*$ is the one that balances the tradeoff between the distortions caused by sampling and transmission. 
\vspace{-1mm}
\subsection{Zero-Wait Sampling Policy}
\begin{figure}[t]
	\centering
	\includegraphics[width=1\columnwidth]{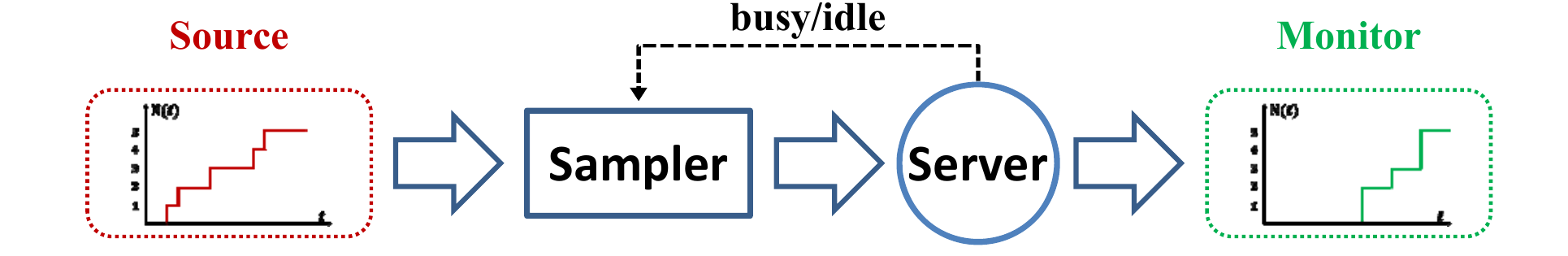}
	\setlength{\abovecaptionskip}{-2mm}
	\caption{The system model for the  zero-wait sampling policy}\label{521508}
	\vspace{-2mm}
\end{figure}

\vspace{-1mm}
In this subsection, we focus on the reconstruction with zero-wait sampling policy. For zero-wait sampling, the sampler only observes the signal when the server is idle, as shown in \figurename\ \ref{521508}. Thus, the sampled packets will be served directly once it is generated, instead of waiting in the queue compared to the case in uniform sampling. 

\begin{figure}[t]
	\centering
	\subfigure[The first few events of sampling and reconstruction: the gap between two curves is the reconstrcuction distortion.]{
		\vspace{-0mm}
		\label{zero_wait_1} 
		\includegraphics[width=0.43\columnwidth]{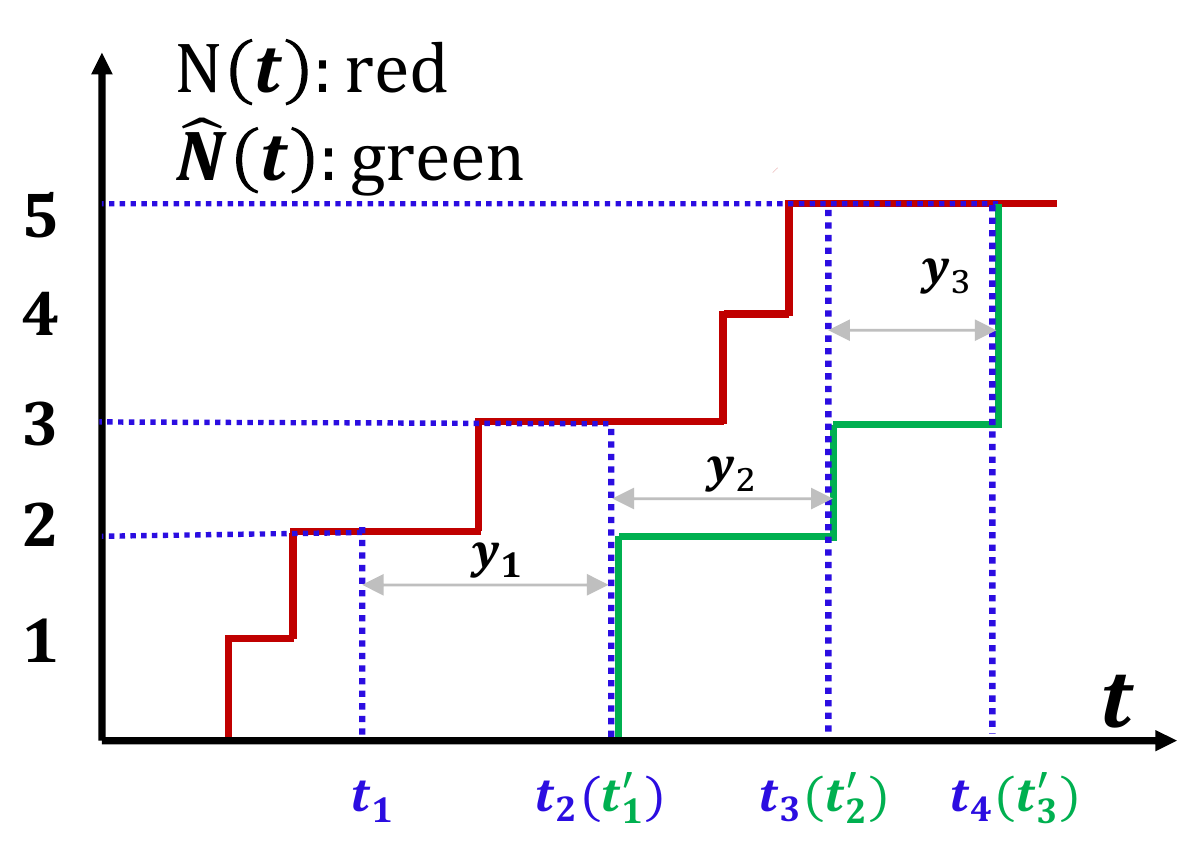}
		\vspace{-0mm}
	}
	\hspace{2mm}
	\subfigure[One of the divided polygons: it can be divided into two sub-polygons.]{
		\vspace{-0mm}
		\label{zero_wait_2} 
		\includegraphics[width=0.48\columnwidth]{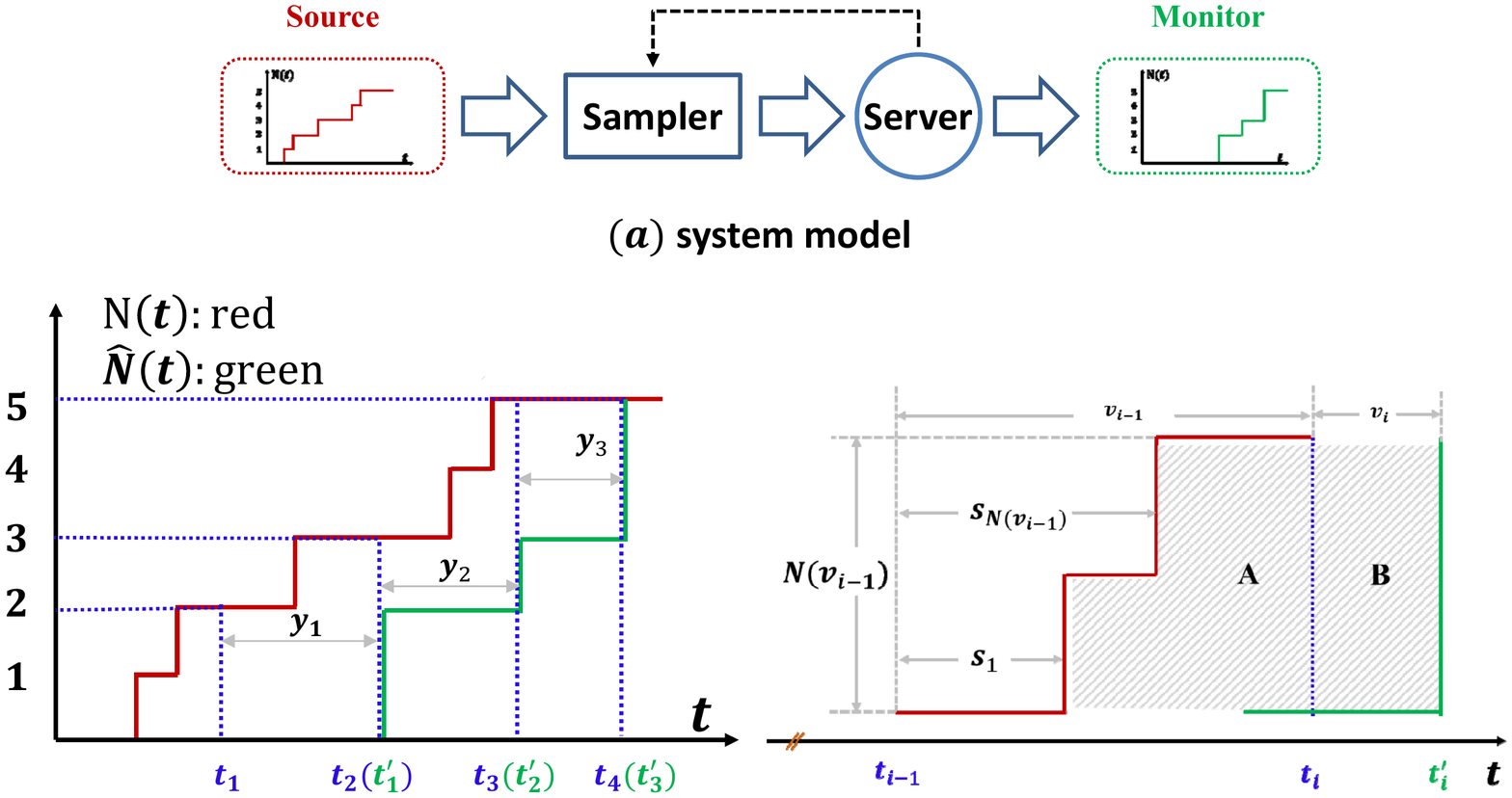}
		\vspace{-0mm}
	}	
	\vspace{-0mm}
	\caption{The sampling and reconstruction under the zero-wait sampling policy}
	\label{10261432} 
	\vspace{-0.2cm}
\end{figure}

 In \figurename\ \ref{zero_wait_1}, we give the first events of the sampling and reconstruction under zero-wait policy.  The overall distortion area can be divided into polygons, one of which is shown in \figurename\ \ref{zero_wait_2}\footnote{Since sample $N(t_i)$ is obtained at $t_i$ immediately after the previous sample is served, the sampling interval $(t_{i+1}-t_{i})$, the service time of sample $N(t_i)$, i.e., $v_i$ in \figurename\ \ref{zero_wait_2}, and the  inter-arrival time of samples $N(t_i)$ and $N(t_{i+1})$, i.e., $y_i$ in \figurename\ \ref{zero_wait_1}, are exactly same and can be captured by $i.i.d.$ exponential random variables, whose $p.d.f$ are given in \eqref{11121516}.}. In detail, the polygon is circled by curve  $\{N(t), t_{i-1}\leqslant t \leqslant t_i\}$, line $\left[\left(t_{i-1}, N(t_{i-1})\right), (t_i^{'}, N(t_{i-1}))\right]$, line $\left[(t_i^{'}, N(t_{i-1})),(t_i^{'}, N(t_{i}))\right]$, and line $\left[(t_i, N(t_{i})),(t_i^{'}, N(t_{i}))\right]$. 
 \vspace{-0mm}
 
 \begin{lemma}\label{lemma5}
 	The average distortion under zero-wait sampling policy is denoted by $\Theta_{zw}$ and can be calculated as
 	 \vspace{-0mm}
 	 \begin{align}
 	\Theta_{zw} = \mu \mathbb{E}\left\{S_{\Delta}\right\},\label{12021102}
 	\end{align}
 	where $\mathbb{E}\left\{S_{\Delta}\right\}$ is the average area of the divided polygons.
 \end{lemma}
 \vspace{-1mm}
\begin{IEEEproof}
	Based on \eqref{5101505}, $\Theta_{zw}$ can be reformulated as
	 \begin{align}
	\Theta_{zw} &= \frac{1}{T}\int_0^T D(t){\rm{d}}t = \frac{I}{T}\frac{1}{I}\sum\limits_{i = 1}^{I}\Delta_i.\label{521705}
	\end{align}
	Recall the division of the polygons, the total number of the polygons $I$ equals the total sampling times as well as the total transmission times. Thus, the first item in \eqref{521705} equals $\mu$. The term $\sum\nolimits_{i = 1}^{I}\Delta_i$ is the average area of the polygons. Thus, we arrive at \eqref{12021102}.
\end{IEEEproof}
 \vspace{-0mm}
\begin{lemma}\label{lemma6}
	The average area of the polygons is given as
	\begin{align}\label{1221223}
	\mathbb{E}\left\{S_{\Delta}\right\} = \frac{2\lambda}{\mu^2}.
	\end{align}
\end{lemma}
 \begin{IEEEproof}
	As shown in \figurename\ \ref{zero_wait_2}, a polygon $\Delta_i$ can be divided int sub-polygons marked as $A$ and $B$, that account for the sampling distortion and the transmission distortion, respectively. Thus, we have $\mathbb{E}\left\{S_{\Delta}\right\} = \mathbb{E}\left\{S_A\right\} + \mathbb{E}\left\{S_B\right\}$. In \emph{Appendix} \ref{12021125}, we show that $\mathbb{E}\left\{S_A\right\} = \mathbb{E}\left\{S_B\right\} = \frac{\lambda}{\mu^2}$. Thus, we obtain \eqref{1221223}. 
\end{IEEEproof}
 
\begin{theorem}
	The average distortion under zero-wait policy is given as
	\begin{align}\label{5301530}
	\Theta_{zw} = \lambda\frac{2}{\mu}.
	\end{align}
\end{theorem}
 \vspace{-2mm}
\begin{IEEEproof}
Combining the results in \eqref{12021102} and \eqref{1221223}, we obtain the result in \eqref{5301530}.	
\end{IEEEproof}

Since we are calculating the average distortion for a fixed scheme, i.e., zero-wait sampling, the average distortion is a constant value, given the arrival rate of the original process $\lambda$ and the server capacity $\mu$. Zero-wait sampling policy is simple and easy to apply since there is no need to consider the buffer capacity of the server which is quite limit for portable devices.

 \vspace{-0mm}
\section{Simulation Verification}
 \vspace{-0mm}
In this section, we present simulation results to validate the theoretical expressions of the derived average distortion under various sampling policies as well as the performance of the proposed algorithm. The simulation horizon $T$ is set as $10^6$.  
 \vspace{-1mm}
\subsection{The Uniform Sampling Policy}
 \vspace{-1mm}
\begin{figure}[t]
	\centering
	\subfigure[The average distortion and the average queueing delay]{
		\label{fig:subfig:a} 
		\includegraphics[width=0.62\columnwidth]{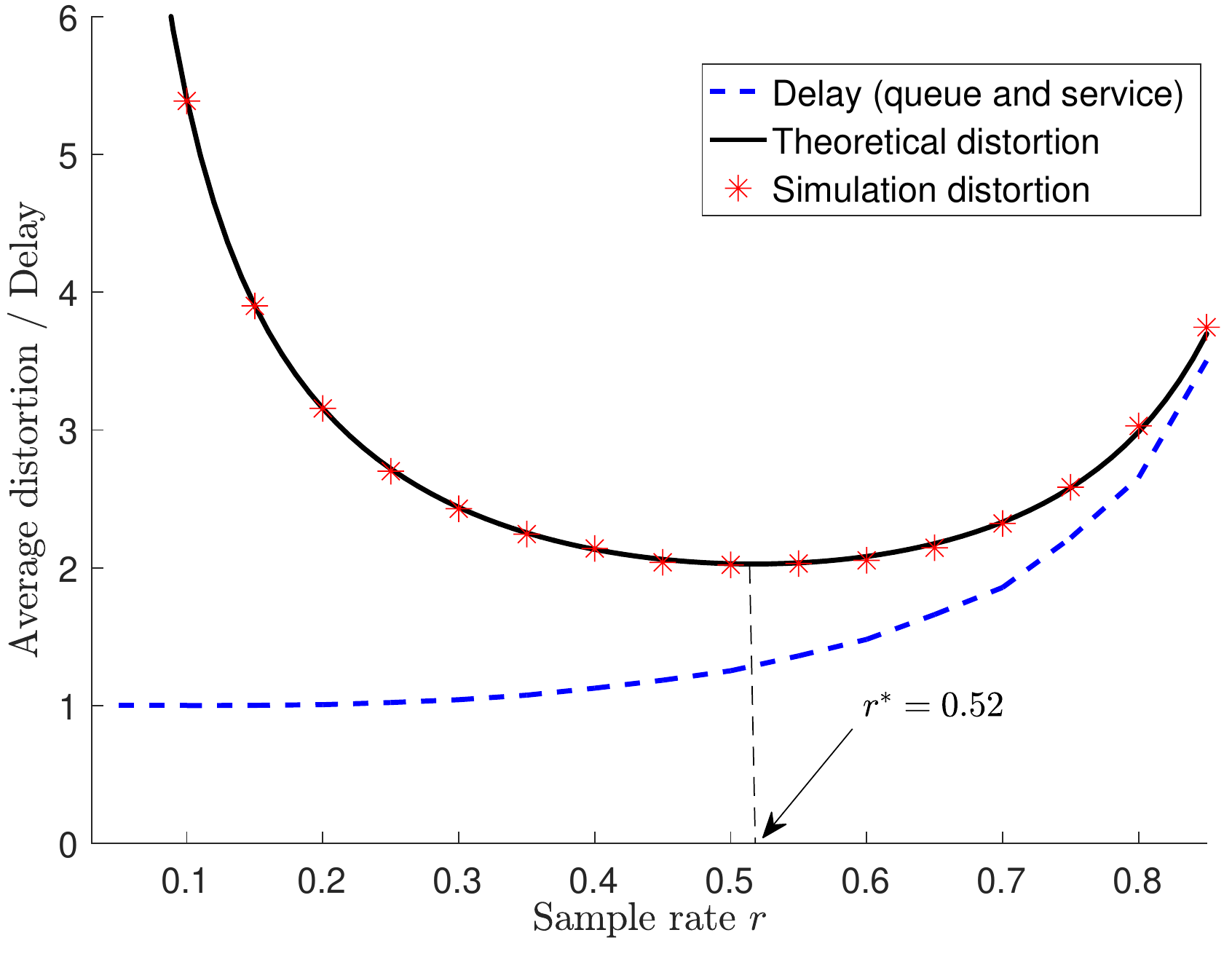}}\\
	\subfigure[The average distortion and the age of information]{
		\label{fig:subfig:b} 
		\includegraphics[width=0.65\columnwidth]{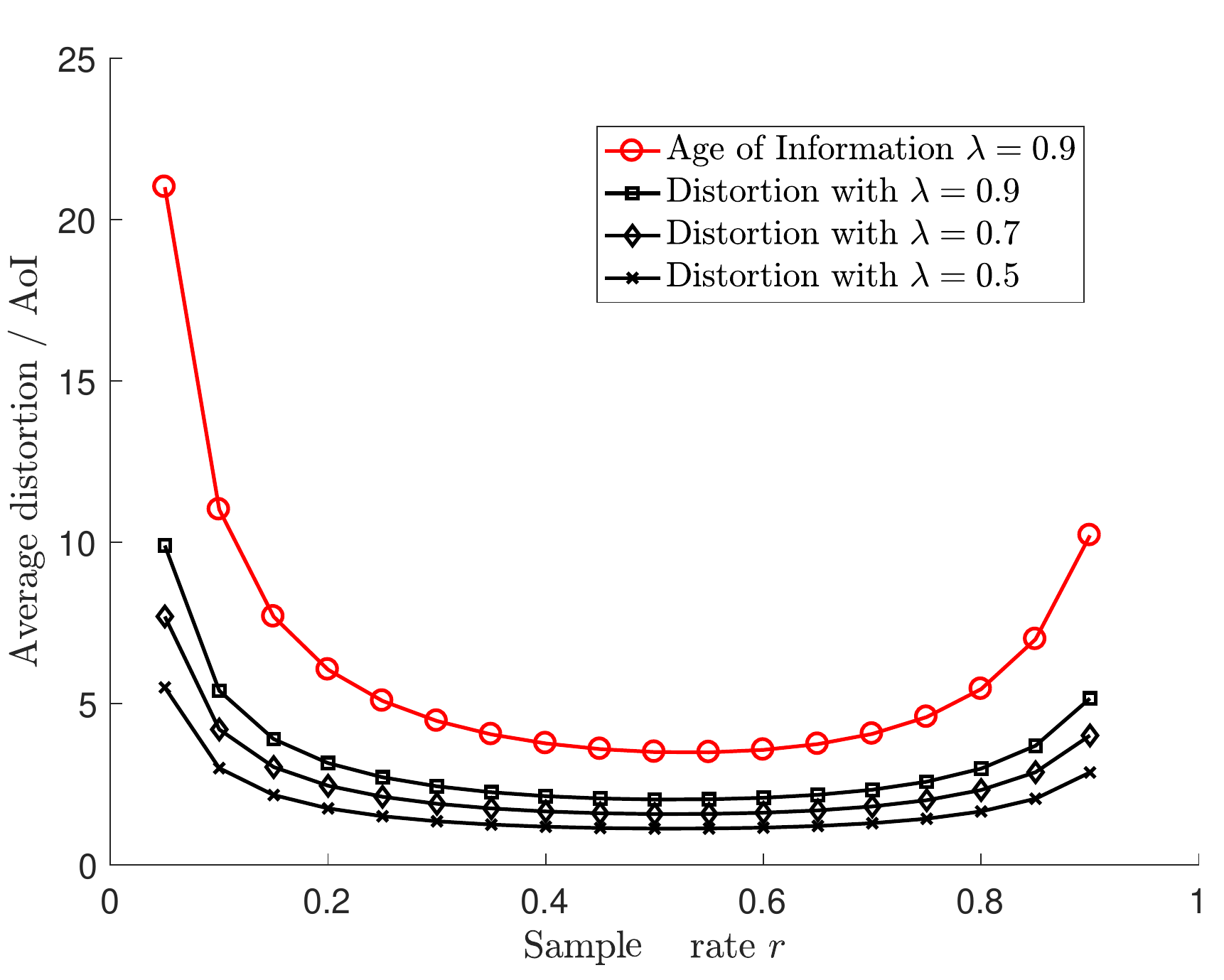}}
	\caption{Simulation results for the uniform sampling policy: the arrival rate of $N(t)$ for \figurename\ \ref{fig:subfig:a} is set as $\lambda=0.9$ and the average service rate for both \figurename\ \ref{fig:subfig:a} and \figurename\ \ref{fig:subfig:b}  is set as $\mu = 1$. }
	\label{1301752} 
	\vspace{-6mm}
\end{figure}

In \figurename\ \ref{1301752}, we present the simulation and theoretical results for the uniform sampling method. In \figurename\ \ref{fig:subfig:a}, we first plot the theoretical and the simulation results of the average distortion, as given in the black line and the line marked with red stars, respectively. We can see that, the simulation results are in accord with the theoretical results, which validates the accuracy of the average distortion-sampling rate function  $\Theta(r)$ in \eqref{3131106}. What's more, the tendency of the curve fits our expectation, i.e., it first decreases as the sampling rate increases due to more samples of the original signal are available by the receiver for the reconstruction, then increases due to that plentiful samples cause high waiting delay of the queueing system. Second, we also plot the average delay (the queueing delay plus the service time) experienced by the samples as given in the blue dash line. We can see that, the average delay increases as the sampling rate increases and the increasing is dramatical when the sampling rate is high. As a result, this dramatical increase causes the increase of average distortion. It shows that the defined average distortion reflects not only the traditional sampling distortion but also the transmission delay. It is a result of the tradeoff between the sampling and the transmission during the reconstruction.

Once we validate the correctness of the average distortion-sampling rate function, it can be used to find the optimal sampling rate which brings the minimum average distortion. For the adopted parameters in \figurename\ \ref{fig:subfig:a}, the optimal sampling rate is obtained approximately as $0.52$, which means that if we sample with this rate, the receiver can construct the best approximate of ${N}(t)$ in average. 

In \figurename\ \ref{fig:subfig:b}, we compare the average distortion and the age of information (AoI). The similarity is that they show the similar tendency when the sampling rate increases, i.e., they both first decrease and then increase. This is due to that these two metrics both take the sampling and the transmission behaviors into account. However, they are two different metrics. The AoI is only related to the sampling rate, which means that the AoI is a fix value for different processes. In our case, it means for counting process with different arrival rates $\lambda$, they have same AoI. Apparently, it is not enough to evaluate the freshness of the received message. In \figurename\ \ref{fig:subfig:b}, we also give the average distortions for counting processes with different arrival rates. For a given sampling rate, the process with greater arrival rate will induces higher average distortion, which reflects the properties of the original process.

\vspace{-1mm}
\subsection{Reconstruction with Interpolation}
\vspace{-1mm}

\begin{figure}[t]
	\centering
	\includegraphics[width=0.65\columnwidth]{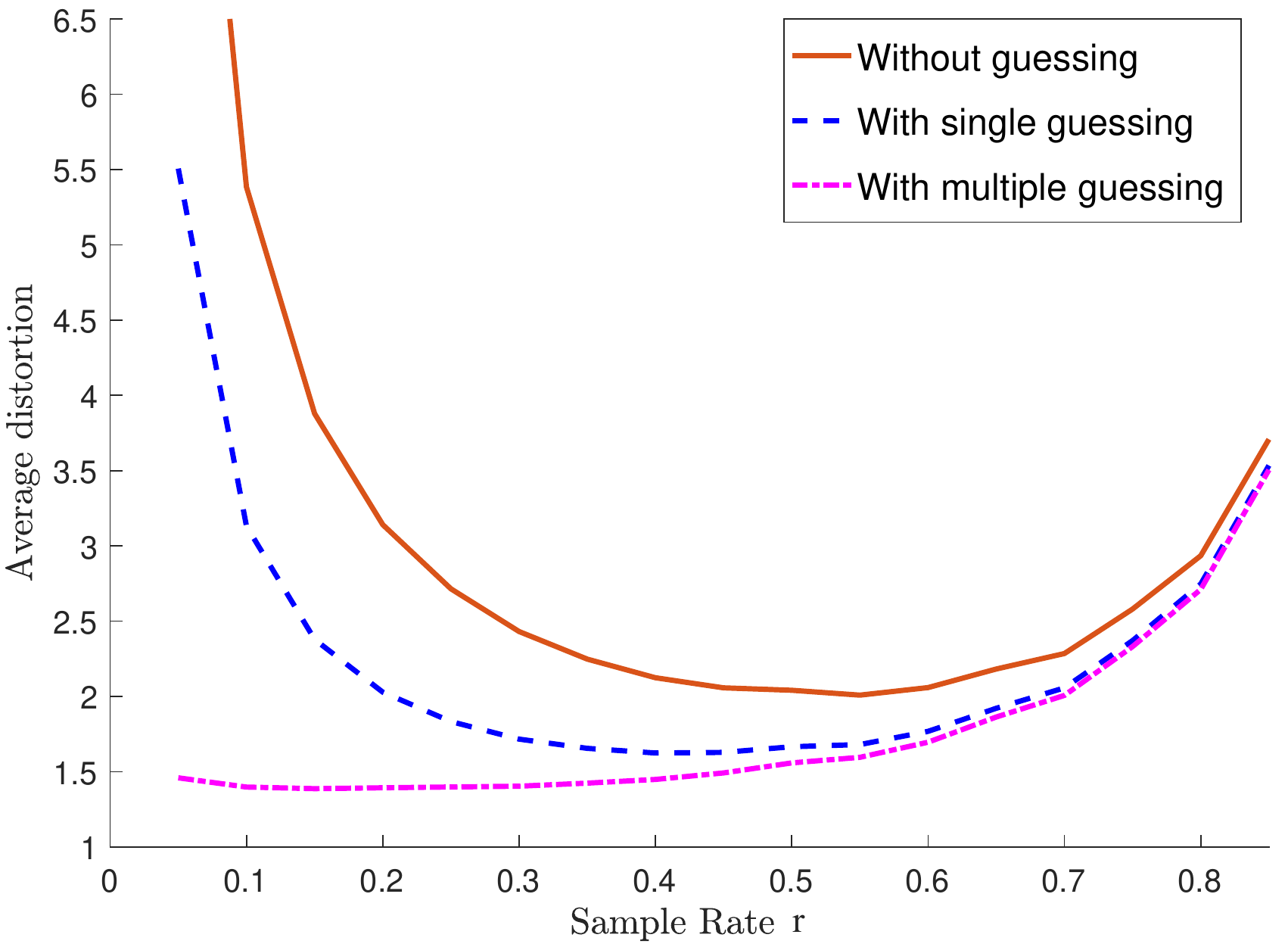}
	\setlength{\abovecaptionskip}{-2mm}
	\caption{The simulation results for the proposed algorithm: the monitor is able to make some guesses for the omitted details based the previously received samples.}\label{4241722}
	\vspace{-2mm}
\end{figure}

In \figurename\ \ref{4241722}, we give the simulation results of the proposed algorithm. The simulation parameters are set as same as the parameters in \figurename\ \ref{1301752}. The yellow solid line is the average distortion when the reconstruction is based only on the received samples. We can see that, for smaller sampling rate, the average distortion is very high since many details of the counting process are omitted and just finite snapshots of the original process are available to the monitor. The blue dash line is the average distortion obtained when only one guessing point is added to the reconstructed process. We can see that, the average distortion is decreased compared to the no guessing case, especially for small sampling rate when the sampling distortion contributes more to the average distortion. The purple dash line is the average distortion obtained exactly based on \emph{Algorithm} \ref{4231925}, where multiple guessing points are added and these points are distributed uniformly between two received samples. The average distortion can be further decreased compared to the blue line. With the idea of adding guessing points, the sampling distortion can be decreased but it improves nothing for the transmission distortion. Thus, neither the blue nor purple lines improve the performance for higher sampling rate.

\subsection{The Non-Uniform Sampling Policy}

In this subsection, we provide the simulation results for the considered two non-uniform sampling policies.

\subsubsection{The threshold-based policy}

\begin{figure}[t]
	\centering
	\subfigure[The simulation results of the average distortion for different arrival rates of original process $N(t)$: the service rate $\mu$ is set as 1]{
		\label{thres:subfig:a} 
		\includegraphics[width=0.7\columnwidth]{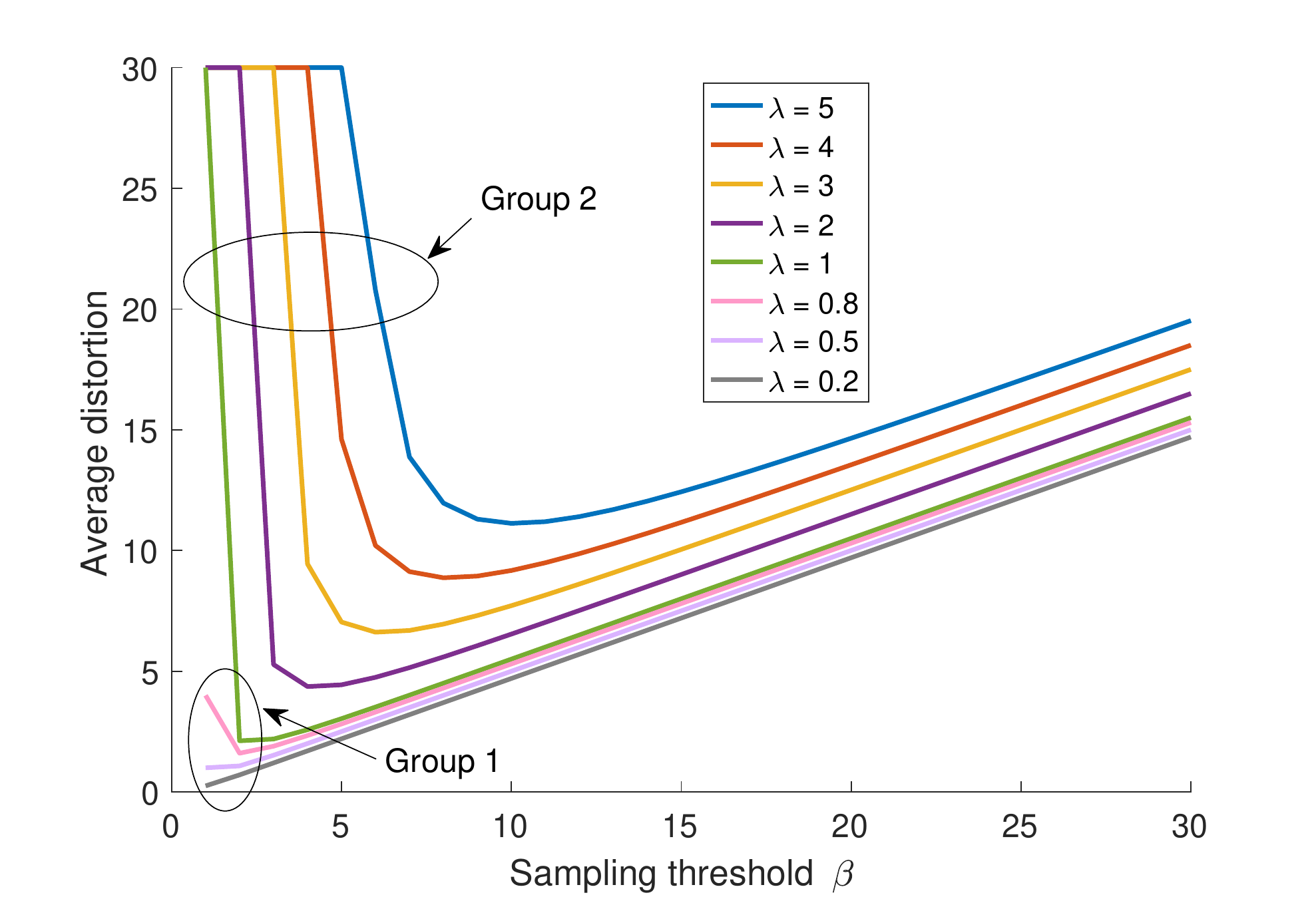}}\\
	\subfigure[The average distortion changes with the service rate]{
		\label{thres:subfig:b} 
		\includegraphics[width=0.48\columnwidth]{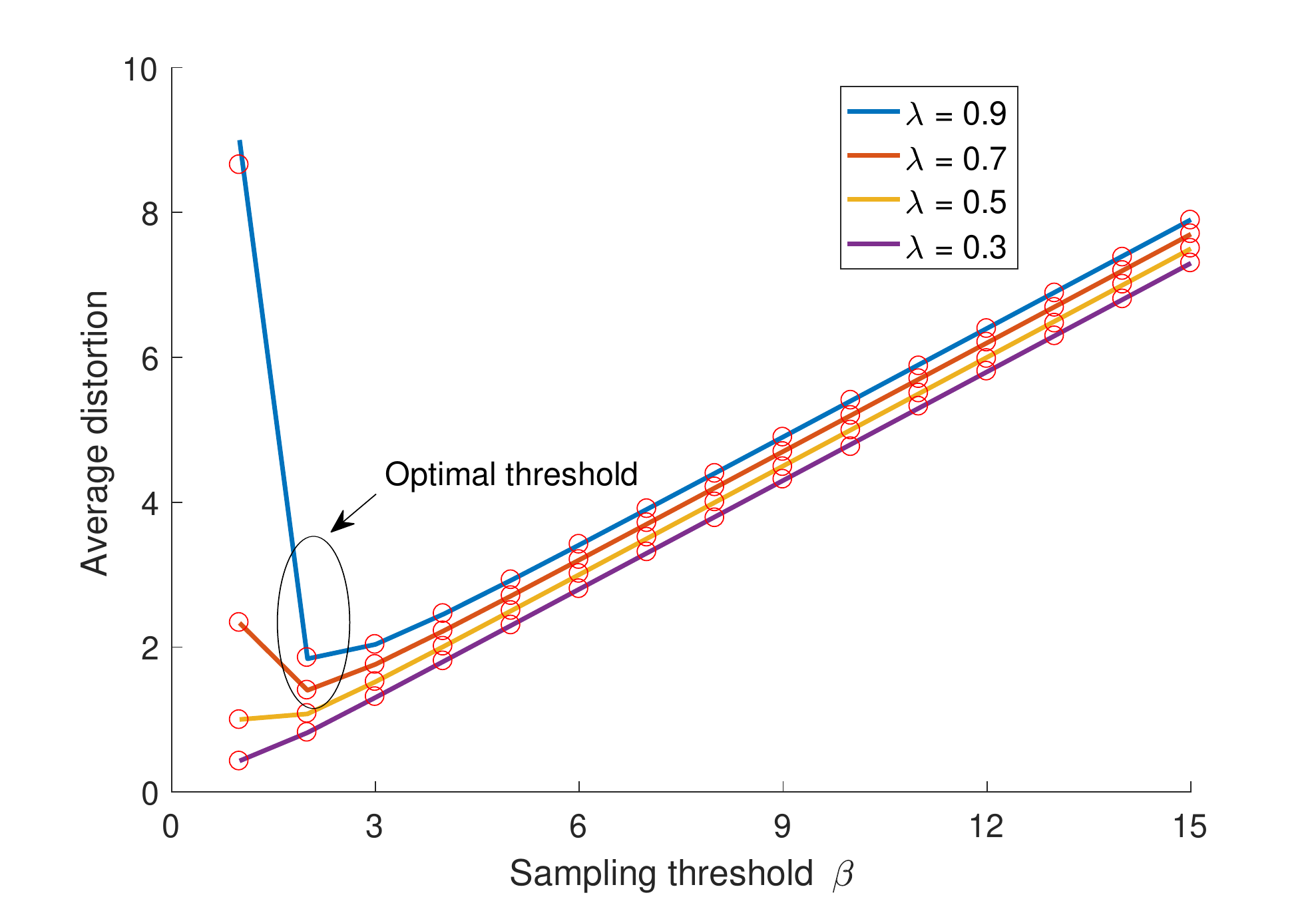}}
	\subfigure[The average distortion changes with the service rate]{
		\label{thres:subfig:c} 
		\includegraphics[width=0.48\columnwidth]{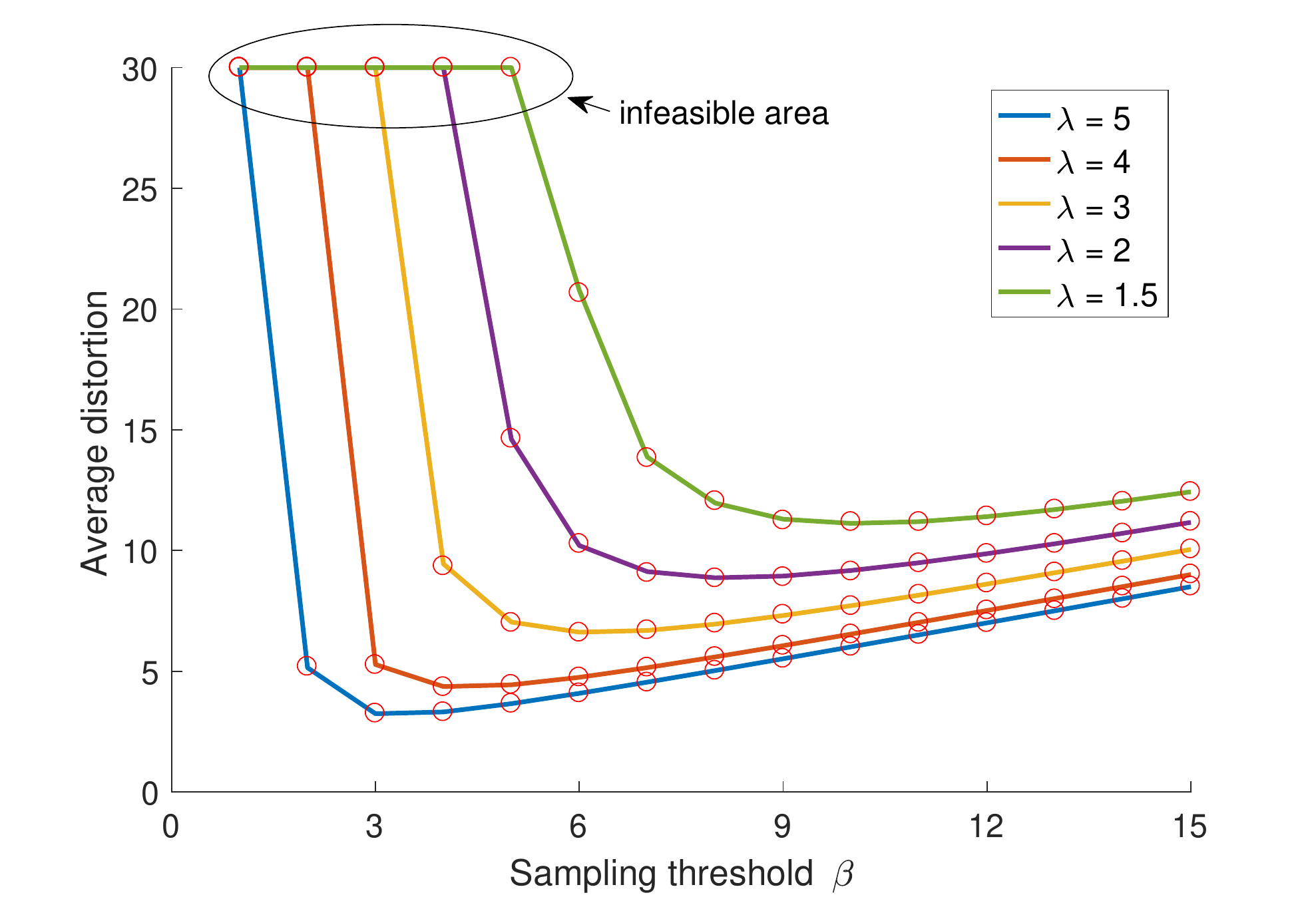}}	
	\caption{The simulation results for the threshold-based sampling policy}
	\label{5302310} 
\end{figure}

we give the simulation results for the threshold-based sampling method in \figurename\ \ref{5302310}. Firstly, with the service rate $\mu$ being $1$, we give the simulation results of the average distortion for a range of arrival rate $\lambda$ of the original process in \figurename\ \ref{thres:subfig:a}. We can see that, these curves can be divided into two groups as the circles. The first group is when the arrival rate is below than $\mu=1$. In this case, the queueing system is stable and the average distortion is finite even when the threshold is $1$, i.e., any update of $N(t)$ will be recorded and transmitted which brings heavy traffic to the queue. The other group is when the arrival rate $\lambda \geqslant 1$. In this case, the heavy traffic bought by the small threshold makes the queue system unstable and leads to infinite average distortion. We set a great number (it is $30$ in the figure) if the queue is unstable. With the threshold increasing, the queueing system becomes stable. 

\begin{figure}[t]
	\centering
	\subfigure[The average distortion changes with the arrival rate ]{
		\label{zero_wait:subfig:a} 
		\includegraphics[width=0.65\columnwidth]{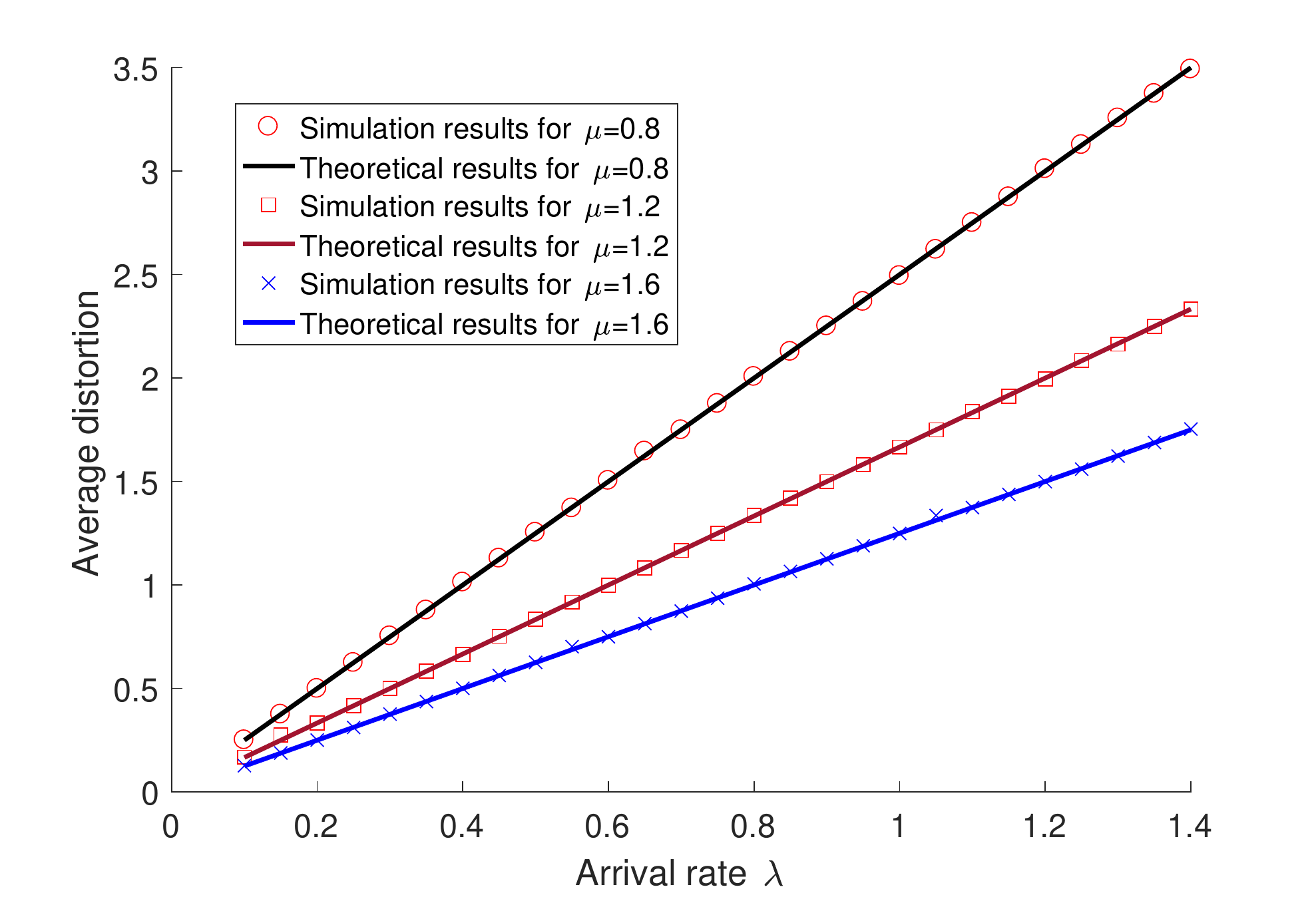}}\\
	\subfigure[The average distortion changes with the service rate]{
		\label{zero_wait:subfig:b} 
		\includegraphics[width=0.65\columnwidth]{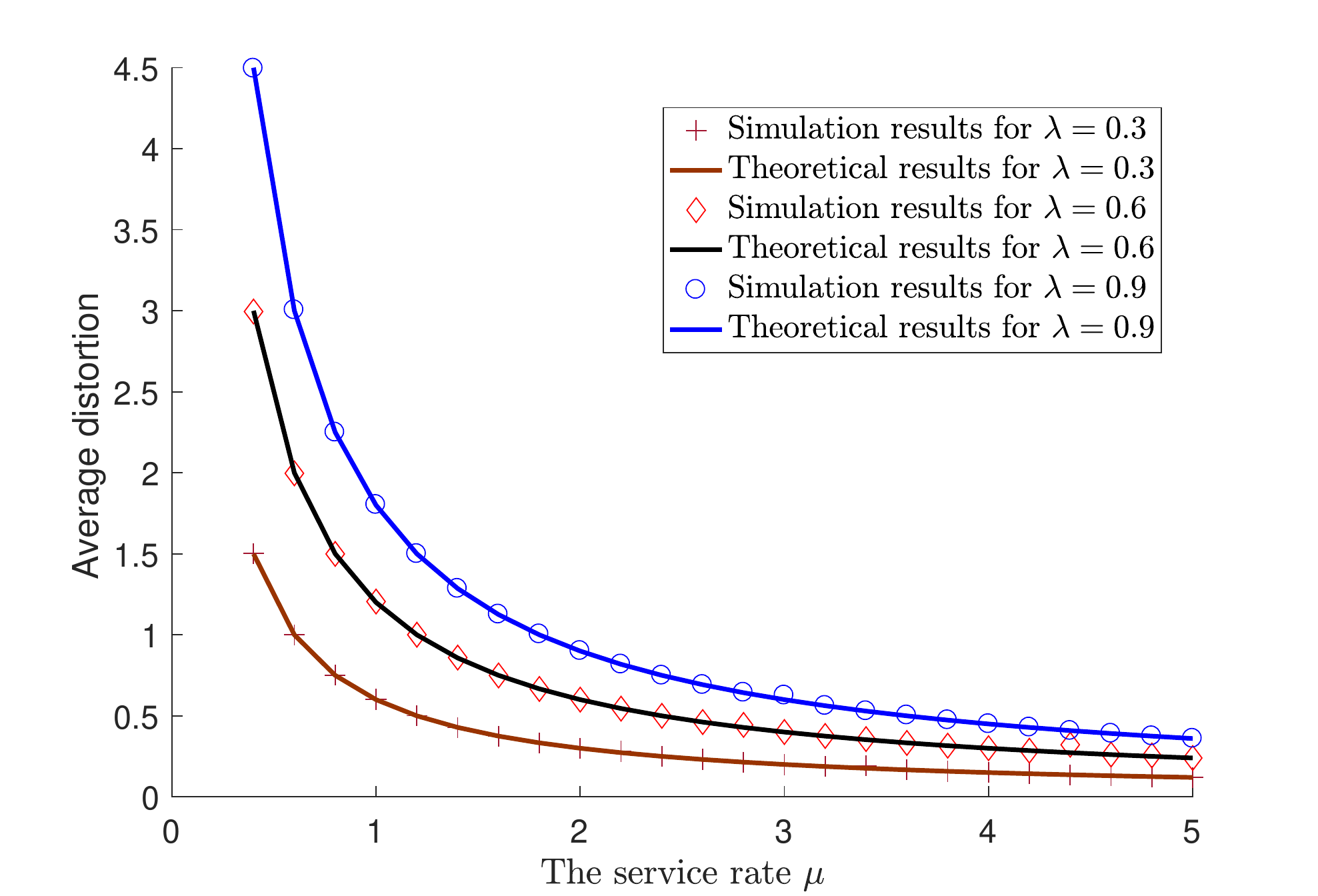}}
	\caption{Simulation results for the zero-wait sampling policy}
	\label{5301521} 
	\vspace{-2mm}
\end{figure}

The theoretical and simulation results for the two groups are given in \figurename\ \ref{thres:subfig:b} and \figurename\ \ref{thres:subfig:c}, respectively. The accordance validates the expression of the average distortion given in \eqref{10281146}. In \figurename\ \ref{thres:subfig:b}, increasing the threshold $\beta$ when sampling may not bring any benefits for smaller $\lambda$, as shown in the cases $\lambda=0.3$ and $\lambda=0.5$. For greater arrival rates, such as the cases  $\lambda=0.7$ and $\lambda=0.9$, the systems works in a relative high load mode for threshold $\beta=1$. Thus, increasing the threshold a little bit greater to $\beta=2$ will decrease the average distortion distinctly. Thus, the optimal threshold $\beta^*=1$ for smaller $\lambda$ and $\beta^*=2$ for greater $\lambda$. In \figurename\ \ref{thres:subfig:c}, there exist infeasible thresholds for all cases as shown in the left upper area of the figure. For this group, the optimal thresholds vary for different arrival rates. Using the theoretical average distortion, it is easily to derive the optimal threshold value as $3$, $4$, $6$, $8$, and $10$ for $\lambda = 1.5$, $2$, $3$, $4$, and $5$, respectively.

\subsubsection{The Zero-Wait Sampling Policy}

we give the simulation and theoretical results of the average distortion under the zero-wait sampling policy in \figurename\ \ref{5301521}. Considering that the zero-wait policy is a fix policy, the average distortion is a constant value when the arrival rate of $N(t)$ and the service rate are given. We plot the distortion-arrival rate and the distortion-service rate curves in \figurename\ \ref{zero_wait:subfig:a} and \figurename\ \ref{zero_wait:subfig:b}, respectively. In both cases, the correctness of the average distortion given in \eqref{5301530} can be validated, since the simulation and the theoretical results are in agreement. Specifically, in \figurename\ \ref{zero_wait:subfig:a}, for a fixed service rate, the average distortion increases linearly with the increasing of arrival rate of $N(t)$. In \figurename\ \ref{zero_wait:subfig:b}, for a fixed arrival rate, the average distortion decreases inverse proportionally with the increasing of the service rate $\mu$. 

\section{Conclusion}

In this paper, we consider the real-time remote reconstruction of a counting process, where distortion occurs due to the sampling and the queueing behaviors. The average distortion defined as the average gap between the original signal and the reconstructed signal is adopted to measure the performance of the reconstruction. We study the uniform sampling method and two non-uniform sampling policies, i.e., the threshold-based policy and the zero-wait policy. For all the three policies, we derive the theoretical expressions of the average distortion as functions of some key sampling parameters. The theoretical results help us find the optimal sampling parameters that balance the tradeoff between the sampling distortion and the transmission distortion, and then induce the minimum average distortion. Simulation results validate our derivation and at the mean time provide a way to understand the defined average distortion. Future works include pursuing theoretical results for other counting processes with other sampling method and packet scheduling policies.

\appendices

\section{The finite packet length of the sample}\label{12011339}

For uniform sampling, samples are obtained at time instants $\{t_i = id, i\geqslant 1\}$. It means that one sample should be able convey the increase of the original process during one sample interval $d$, i.e., $N(d)$. For the Poisson process, we know that
\begin{align}
\text{Pr}\{N(d)=n\} = \frac{(\lambda d)^n}{n!}e^{-\lambda d}, \quad n \in \mathbb{N}.
\end{align} 
Suppose that the length of one packet that used to convey the raw information is $k$ bits. Then, the effective range of one packet is $\{0,1,\cdots,M = 2^k-1\}$. Once $N(d)$ exceeds this range, the sample fails to exactly convey the increase. The probability of failure can be calculated as
\begin{align}
\varrho = \sum\limits_{i=M}^{\infty} \text{Pr}\{N(d)=n\}.
\end{align}
Failure probability $\varrho$ decreases with the increase of $k$. However, a greater $k$ will lead to waste, especially when $N(d)$ is small. In this work, we ideally assume that any $N(d)$ can be exactly expressed by one sample.

\vspace{-0mm}
\section{The calculation of $\mathbb{E}\{S_A\}$, $\mathbb{E}\{S_B\}$, and $\mathbb{E}\{S_C\}$ in \eqref{12011545}} \label{12011542}

In this section, we given $\mathbb{E}\{S_A\}$ in \emph{Lemma} \ref{lemma7},  $\mathbb{E}\{S_B\}$ and $\mathbb{E}\{S_C\}$ in \emph{Lemma} \ref{lemma8}.
\vspace{-3mm}
\begin{lemma}\label{lemma7}
	Area $S_A$ is the distortion caused by sampling and its average is given as 
	\begin{align}\label{11121110}
	\mathbb{E}\{S_A\}  = \frac{\lambda}{2r^2}. 
	\end{align}
\end{lemma}
\vspace{-3mm}
\begin{IEEEproof}
	Let $N(d)$ denote the number of the events that happen during sampling interval $d$, i.e., the height of the polygon, we can obtain $S_A$ as 
	\begin{align}\label{521630}
	S_A = \sum\limits_{k=1}^{N(d)}(d-s_{k}).
	\end{align}
	Clearly, the average area of $A$, i.e., $\mathbb{E}\{S_A\}$, is dependent on the stochastic property of the original counting process $N(t)$. For Poisson counting process, the heights of $\Delta_i$, $i \geqslant 1$ are $i.i.d$ stochastic Poisson variables with parameter $\lambda d$. In order to calculate $\mathbb{E}\{S_A\}$, we first calculate the conditional average when $N(d)$ is given as\footnote{We define $\{Y_{k}, 1\leqslant k \leqslant n\}$ are independent random variables that are uniformly distributed on interval $(0,d)$, see \emph{Theorem} \ref{4242019} in \emph{Appendix} \ref{11121349}.}
	\begin{align}\label{10291500}
	\mathbb{E}\{S_A|N(d) = n\}& \overset{\textcircled{\scriptsize 1}}{=} nd - \mathbb{E}\left\{\sum\nolimits_{k=1}^{n}s_k | N(d) = n\right\} \\
	&\overset{\textcircled{\scriptsize 2}}{=}  nd - \mathbb{E}\left\{\sum\nolimits_{k=1}^{n}Y_{k}\right\}  \nonumber \\
	& \ {=}  nd - \frac{nd}{2} = \frac{nd}{2}, \nonumber
	\end{align}
	where step $\textcircled{\scriptsize 1}$ holds because of the condition $N(d)=n$ and step $\textcircled{\scriptsize 2}$ holds due to \emph{Theorem} \ref{4242019} in  \emph{Appendix} \ref{11121349}. Thus, we can obtain $\mathbb{E}\{S_A\}$ as
	\begin{align}\label{1291115}
	\mathbb{E}\{S_A\} & = \mathbb{E}\left\{\mathbb{E}\left\{S_A|N(d) = n\right\}\right\} \nonumber\\
	& = \sum\nolimits_{n=0}^{\infty} \text{Pr}\{N(d) = n\} \mathbb{E}\{S_A|N(d) = n\} \\
	& = \sum\nolimits_{n=0}^{\infty} \text{Pr}\{N(d) = n\} \frac{nd}{2} \nonumber\\
	& = \frac{d}{2}\mathbb{E}\{N(d)\} =  \frac{\lambda d^2}{2}. \nonumber
	\end{align}
	Considering the relationship between $\!d$ and $\!r$, i.e., $\!$$r\!=\!\frac{1}{d}$, we arrive at the conclusion \mbox{in \eqref{11121110}}.
	
\end{IEEEproof}

\vspace{-4mm}

\begin{lemma}\label{lemma8}
	The sum of the average areas of sub-polygons $B$ and $C$ is given as
	\begin{align}\label{11121140}
	\mathbb{E}\{S_B + S_C\} = \frac{\lambda}{r} \left( \frac{\sigma}{\mu(1-\sigma)} + \frac{1}{\mu} \right) .
	\end{align}
\end{lemma}

\begin{IEEEproof}
	Sub-polygons $B$ and $C$ are rectangles, whose areas can be calculated as $
	S_B = N(d)w_i$ and $S_C = N(d)v_i$, respectively, where $w_i$ is the queueing delay and $v_i$ is the service time of the $i$th sample packet. With uniform sampling, the queueing model is formulated as a $D/M/1$ model. Thus the average queueing delay and the average service time have already been calculated \cite{gross2008fundamentals} and given as $\mathbb{E}\{w_i\} = \frac{\sigma}{\mu(1-\sigma)}$ and $\mathbb{E}\{v_i\} = \frac{1}{\mu}$, respectively, where the parameter $\sigma$ is the solution of Lambert W function given in \eqref{12011602}. Given sampling rate $r$, $N(d)$ is independent with $w_i$ and $v_i$. Thus, we obtain 
	\begin{align}
	\mathbb{E}\left\{S_B + S_C\right\}  = \mathbb{E}\left\{N(d)\right\}\Big( \mathbb{E}\left\{w_i\right\} + \mathbb{E}\left\{v_i\right\} \Big) 
	 = \lambda d \left( \frac{\sigma}{\mu(1-\sigma)} + \frac{1}{\mu} \right).
	 \label{4241049} 
	\end{align}
	With $r=\frac{1}{d}$, we arrive at the conclusion in \eqref{11121140}.
\end{IEEEproof}

\vspace{-0mm}
\section{The Proof of Step $\textcircled{\scriptsize 2}$ in \eqref{10291500}} \label{11121349}
\vspace{-0mm}
	In \emph{Theorem} \ref{4242019}, we give the detailed proof of Step $\textcircled{\scriptsize 2}$ in \eqref{10291500}.
\vspace{-2mm}
	\begin{theorem}\label{4242019}
		Given $N(d) = n$, $\{Y_{k}, 1\leqslant k \leqslant n\}$ are independent random variables that are uniformly distributed on interval $(0,d)$, we have 
		\begin{align}
		&\mathbb{E}\left\{\sum\limits_{k=1}^{n}s_k \Big| N(d) = n\right\} =  \mathbb{E}\left\{\sum\limits_{k=1}^{n}Y_{(k)}\right\} = \frac{nd}{2}. \label{4242018}
		\end{align}
	\end{theorem}
	
	\begin{IEEEproof}
		Considering $s_k$ is the occur times of the $k$th event and given $N(d) = n$ means there already have $n$ events occurred, then the occur time of these $n$ packets must meet the requirement that $s_1 \leqslant s_2 \leqslant \cdots \leqslant s_k  \leqslant \cdots \leqslant s_n$. Based on the conclusion for Poisson process that given $N(d) = n$, the occur times $s_k,\ 1\leqslant k \leqslant n$ have the same distribution of $Y_{(k)}$s, where $Y_{(k)}$ is the $k$th smallest value among $Y_{k}$\cite{ross1996stochastic}. Thus, we have
		\begin{align}
		\mathbb{E}\left\{\sum\limits_{k=1}^{n}s_k \Big| N(d) = n\right\} = \mathbb{E}\left\{\sum\limits_{k=1}^{n}Y_{(k)}\right\}.\label{4242102}
		\end{align}
		In the right hand of \eqref{4242102}, the sequence of variables do not matter since we are calculating the sum. Thus, we have
		\vspace{-2mm}
		\begin{align}
		\mathbb{E}\left\{\sum\limits_{k=1}^{n}Y_{(k)}\right\} = \mathbb{E}\left\{\sum\limits_{k=1}^{n}Y_{k}\right\}.\label{4242107}
		\end{align}
		\vspace{-2mm}
		Remember that, $Y_{k}$ is a random variable that uniformly distributed on interval $(0,d)$, we have $\mathbb{E}[Y_k] = \frac{d}{2}$. Thus, 
		\begin{align}
		\mathbb{E}\left\{\sum\limits_{k=1}^{n}Y_{k}\right\} =\sum\limits_{k=1}^{n}\mathbb{E}\{Y_{k}\} = \frac{nd}{2}.\label{4242110}
		\end{align}
		By far, we have completed the proof.
	\end{IEEEproof}

\section{Detailed calculation for \eqref{10291505} and  \eqref{10291504}} \label{12011714}

In this section, we derive the expressions given in \eqref{10291505} and  \eqref{10291504}.

We first calculate \eqref{10291505} for case $1$. Recall that, $p(x)$ is the $p.d.f.$ of variable $x_i=w_i+v_i$. For case $1$, $t_{i-1}^{'} \geqslant t_i$. Thus, given the condition that there are $n$ events occurring during the sampling interval, the area of the polygon can be drawn as 
\begin{align}
\varphi = \int_d^\infty p(x) \sum\limits_{m=1}^n (x-s_n) {\rm{d}}x.
\end{align}
Combining the result in \emph{Theorem} \ref{4242019}, we know 
\begin{align}
\mathbb{E}[\varphi] = \int_d^\infty p(x) \mathbb{E}\left\{ \sum\limits_{m=1}^n (x-s_n) \right\} {\rm{d}}x = \int_d^\infty p(x) \left( nx-\frac{nd}{2} \right) {\rm{d}}x.
\end{align} 
Thus, $\hat{S}_1$ can be calculated as
\begin{align}
	\hat{S}_1 &= \sum\limits_{n=1}^{\infty} \text{Pr}\{N(d)=n\} \mathbb{E}\{\varphi\} \nonumber\\
	&=\sum\limits_{n=1}^{\infty} n\text{Pr}\{N(d)=n\} \int_d^\infty p(x) (x-\frac{d}{2}) {\rm{d}}x.
\end{align}

We then calculate \eqref{10291504} for case $2$. For case $2$, $t_{i-1}^{'} < t_i$. Thus, given the condition that there are $n$ events occurring during the sampling interval, the area of the polygon can be drawn as 
\begin{align}
\varsigma = \int_0^d p(x) \sum\limits_{m=1}^{n}\text{Pr}\{N(x)=m\}\sum\limits_{k=1}^{m}(x-s_k)  {\rm{d}}x.
\end{align}	
With the conclusion in \emph{Theorem} \ref{4242019}, the expectation of $\varsigma$ is calculated as 
\begin{align}
\mathbb{E}[\varsigma] &= \int_0^d p(x) \sum\limits_{m=1}^{n}\text{Pr}\{N(x)=m\} \mathbb{E} \left\{\sum\limits_{k=1}^{m}(x-s_k) \right\}\rm{d}x \nonumber\\
&=\int_0^d p(x) \sum\limits_{m=1}^{n}\text{Pr}\{N(x)=m\} \frac{mx}{2} \rm{d}x.
\end{align}	
Thus, $\hat{S}_2$ can be calculated as
\begin{align}
\hat{S}_2 &= \sum\limits_{n=1}^{\infty} \text{Pr}\{N(d)=n\} \mathbb{E}\{\varsigma\} \nonumber\\
&= \sum\limits_{n=1}^{\infty}\!\text{Pr}\{N(d)=n\} \int_0^d \frac{x}{2}\sum\limits_{m=1}^{n}m\text{Pr}\{N(x)=m\}  p(x) \rm{d}x.
\end{align}

\vspace{-0mm}
\section{The calculation of $\mathbb{E}\left\{S_A\right\}$ and $\mathbb{E}\left\{S_B\right\}$ in \emph{Lemma} \ref{lemma4}}\label{12012359} 
\vspace{-0mm}

\begin{lemma}\label{11121759}
	The area of sub-polygon $A$ is denoted by $S_A$, whose average is obtained as
	\begin{align}\label{11121735}
	\mathbb{E}\left\{S_A\right\} = \frac{(\beta-1)\beta}{2\lambda}.
	\end{align}
\end{lemma}
\vspace{-3mm}
\begin{IEEEproof}
	We first express the area of sub-polygon $A$ as 
	\begin{align}
	S_A = \sum\limits_{n=1}^{\beta}(s_{\beta}-s_n),
	\end{align}
	where $s_{\beta}$ is the total time that needed for $\beta$ events occurring and $\{s_n$, $n = 1, 2, \cdots, \beta\}$ are the occur time of the $n$th event. For Poisson counting process, we know that variable $s_{\beta}$ submits to an Erlang distribution with stage $\beta$ and mean $\frac{r}{\lambda}$. Its $p.d.f.$ is given as
	\begin{align}
	h(x) = \frac{{\lambda}^{\beta}x^{\beta-1}e^{-\beta x}}{(\beta-1)!}, \ x\geqslant 0.
	\end{align}
	We then calculate the conditional expectation $\mathbb{E}\left\{S_A | s_{\beta} = x\right\}$, i.e., 
	\begin{align}
	\mathbb{E}\left\{S_A | s_{\beta} = x\right\} = \sum\limits_{n=1}^{\beta}(x-s_n) = \sum\limits_{n=1}^{\beta-1}(x-s_n),\label{5111128}
	\end{align}
	where the last equation holds due to the fact that $x = s_{n}$ for the term $n=\beta$. Equation \eqref{5111128} is calculated under the condition that there are $(\beta-1)$ events occur in sequence during interval $[0,x)$. Recall the conclusion in \emph{Theorem} \ref{4242019} and let $d=x$ and $n=\beta-1$, we have the following conclusion
	\vspace{-0mm} 
	\begin{align}
	\mathbb{E}\left\{S_A | s_{\beta} = x\right\} = \frac{(\beta-1)x}{2}.
	\end{align}
	Then, we are able to calculate $\mathbb{E}\left\{S_A\right\}$ as
	\begin{align}
	\mathbb{E}\left\{S_A\right\} &= \int_0^{\infty} \mathbb{E}\left\{S_A | s_{\beta} = x\right\} \ h(x) \rm{d}x \nonumber \\
	&= \int_0^{\infty} \frac{(\beta-1)x}{2}\  h(x) \rm{d}x \nonumber\\
	&= \frac{(\beta-1)\beta}{2\lambda}.  \label{5111138}
	\end{align}	
	Thus, we arrive at the conclusion in \eqref{11121735}.
\end{IEEEproof}

\begin{lemma}\label{11121737}
	The area of sub-polygon $B$ is denoted by $S_B$, whose average is given as
	\begin{align} \label{11121753}
	\mathbb{E}\left\{S_B\right\} = \beta \mathbb{E}\left\{w_i\right\} =\frac{\beta}{\mu(z_0^{\beta}-1)}.
	\end{align}
\end{lemma}

\begin{IEEEproof}
	Sub-polygon $B$ is a rectangle, whose area is calculated as 	
	\begin{align}\label{11121742}
	\mathbb{E}\left\{S_B\right\} = \beta \mathbb{E}\left\{w_i\right\},
	\end{align}
	where $w_i$ is the queueing delay of the packets. To obtain $\mathbb{E}\left\{w_i\right\}$, we turn back to the system model in \figurename\ \ref{5101109}(a), where the queueing system can be modeled as an $E_r/M/1$ queueing system. For such a queueing system, it is not adequate to just   calculate the steady-state probability of the queue states $\pi_k$, namely, the probability that there are $k$ sample packets in the queue. Because the packet arrival process has $\beta$ phases even for the same queue state $'queue\  length= k'$, corresponding to the state that there already have $b$ $(0\leqslant b \leqslant \beta-1)$ events occurred when there are $k$ samples queueing in the buffer. By defining the joint steady-state probability as $\pi_{k,b} = \text{Pr}\{queue ~length= k,\  phase = b\}$ and abbreviating it to $\pi_{k,b} = \text{Pr}\{{ql}= k,\  {ph} = b\}$, we have the following transition probabilities
	\vspace{0mm}
	\begin{equation}
	\left\{ \begin{aligned}
	&\text{Pr}\{{ql} = k,\  {ph} = b+1 \ \big|\  {ql} = k,\  {ph} = b\} = \lambda, \quad\quad k \geqslant 0, \ 0\leqslant b < \beta-1; \nonumber\\
	&\text{Pr}\{{ql} = k+1,\  {ph} = 0 \ \big|\  {ql} = k,\  {ph} = \beta-1\} = \lambda, \quad\quad k \geqslant 0; \nonumber\\
	&\text{Pr}\{{ql} = k-1,\  {ph} = b \ \big|\  {ql} = k,\  {ph} = b\} = \mu, \quad\quad k \geqslant 1, \  0\leqslant b \leqslant \beta-1.
	\end{aligned}
	\right.
	\end{equation}
	\begin{figure}[t]
		\centering
		\includegraphics[width=0.95\columnwidth]{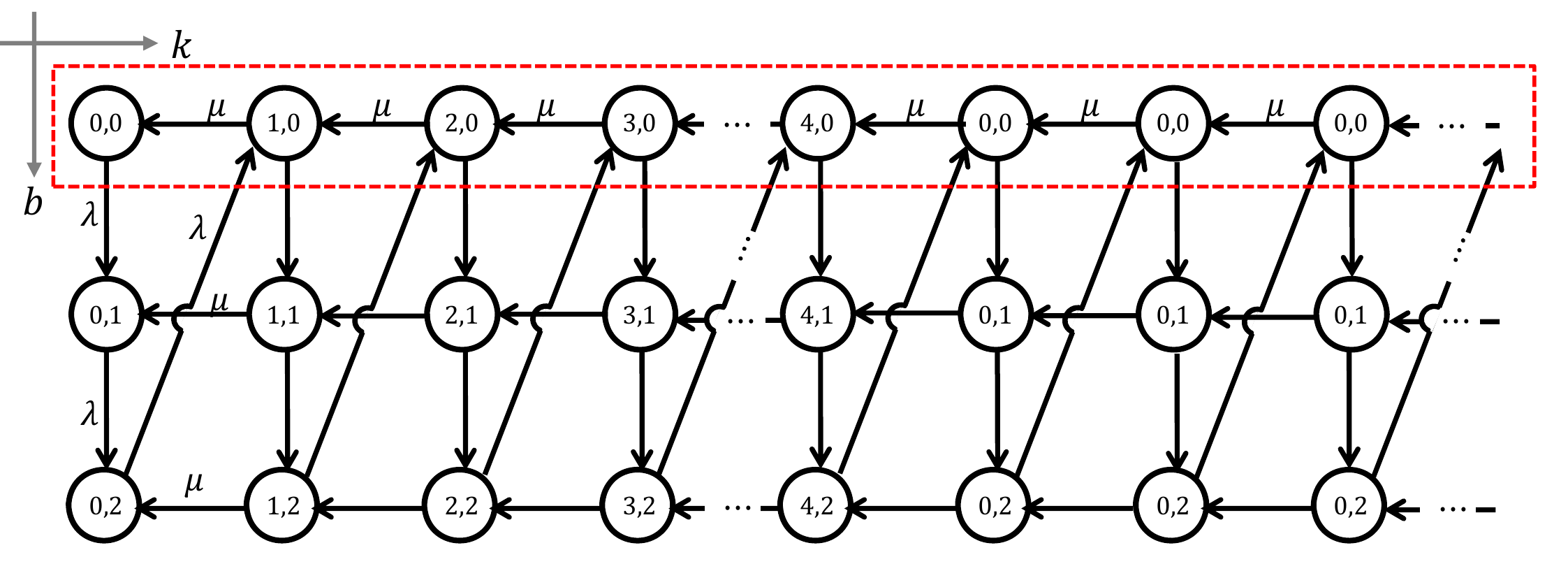}
		\caption{The transitions between joint queue-phase states when the threshold $\beta$ is given as $3$. The horizontal axis is the queue length, i.e., the number of the packets queueing in the buffer while the vertical axis is the arrival phase, i.e., the number of the events that occur since last sampling. The states in the red dash box are the states that observed by new generated sample.   }\label{141441}
		\vspace{-0mm}
	\end{figure}
	
	In \figurename\ \ref{141441}, an example of the transitions between the joint queue-phase states is given with the threshold for the sampling being $3$. As time goes by, one can trace the system state in a zigzag way. For example, starting from state $(0,0)$ which means there is no sample in the queue and no event occurs, the system jumps to state $(0,1)$ with a new event occurring. Since the accumulated change of the original process is less than the threshold $\beta=3$, no sample is generated and the queue state remains as zero. The system then jumps to state $(0,2)$ with another event occurring and still no sample being generated. At state $(0,2)$, one event occurring will increase the phase state to $3$ which  equals to threshold $\beta=3$. Thus, a new sample is generated and enters the queue. This makes the queue state increase to $1$ and the phase state reset to zero, \mbox{and so on.}
	
	However, we do not care all of joint queue-phase states of the system shown in \figurename\ \ref{141441}. Only the states observed by the new generated samples when they enter the queue influence the waiting time, as shown in the red dash box in \figurename\ \ref{141441}. Using the global equilibrium equations and the normalization equation, we can calculate the joint queue-phase steady-state probability $\{\pi_{k,b} \ |\   k \!\geqslant\! 0,\!0 \!\leqslant \!b <\! \beta$\}, among which, the states observed by the new samples are given as 
	\begin{align}
	\pi_{k,0} = \frac{\lambda}{\beta \mu}(z_0-1){z_0}^{-k\beta-1},\  k\geqslant 0.
	\end{align} 
	Since $\{\pi_{k,0}, \ k \geqslant 0\}$ are only part of the whole states, the sum of $\pi_{k,0}$ is less than one. Thus, the steady-state probabilities should be well normalized. Let $\Pi$ be the sum of $\{\pi_{k,0}, \ k \geqslant 0\}$, namely,
	\begin{align}
	\Pi &= \sum\limits_{k=0}^{\infty}\pi_{k,0} = \frac{\lambda(z_0-1)}{\beta \mu z_0 (1-{z_0}^{\beta})}.
	\end{align}
	We then normalize the probabilities as
	\begin{align}
	\pi_{k,0}^* &= \frac{\pi_{k,0}}{\Pi} = {z_0}^{-k\beta}(1-{z_0}^{-\beta}).
	\end{align}
	With $\pi_{k,0}^*$, we can calculate the average queueing length of the queue $\mathbb{E}\left\{q\right\}$ and the average waiting time $\mathbb{E}\left\{w\right\}$ as
	\begin{align}
	\mathbb{E}\left\{q\right\} &= \sum\limits_{k=0}^{\infty}k \pi_{k,0}^* = \frac{1}{z_0^{\beta}-1}, \\ 
	\mathbb{E}\left\{w\right\} &= \frac{\mathbb{E}\left\{q\right\}}{\mu} = \frac{1}{\mu(z_0^{\beta}-1)}, \label{11121752}
	\end{align}  
	respectively. Replacing $\mathbb{E}\left\{w\right\}$ in \eqref{11121742} with the result in  \eqref{11121752}, we arrive at  \eqref{11121753}.
\end{IEEEproof}

\section{The calculation of $\mathbb{E}\left\{S_A\right\}$ and $\mathbb{E}\left\{S_B\right\}$ in \emph{Lemma} \ref{lemma6}}\label{12021125}

\begin{lemma}\label{11121608}
	The area of $A$ is denoted by $S_A$ and the average of $S_A$ is given as
	\begin{align}\label{11121600}
	\mathbb{E}\left\{S_A\right\} = \frac{\lambda}{\mu^2}.
	\end{align}
\end{lemma} 
\begin{IEEEproof}
	The area of sub-polygon $A$ can be obtained as 
	\begin{align}\label{521631}
	S_A = \sum\limits_{k = 1}^{N(v_{i-1})} \left(v_{i-1} - s_k\right),
	\end{align}
	where $v_{i-1}$ is the service time of sample obtained at $t_{i-1}$ and $N(v_{i-1})$ is the number of events occurring during time span $v_{i-1} = t_{i} - t_{i-1}$. In order to calculate the expectation of $S_A$, we first calculate
	\begin{align}\label{521644}
	\mathbb{E}\left\{S_A|v_{i-1} = d\right\} = \mathbb{E}\left\{\sum\limits_{k = 1}^{N(d)} \left(d - s_k\right)\right\}.
	\end{align}
	Note that, the part in the bracket of the right hand of \eqref{521644} is exactly the same as  \eqref{521630}. Thus, we can use the conclusion obtained in \eqref{1291115}, i.e,
	\begin{align}
	\mathbb{E}\left\{S_A|v_{i-1} = d\right\} = \frac{\lambda d^2}{2}.
	\end{align}
	
	Remember that, $v_{i-1}$ is an exponential random variable with parameter $\mu$. Thus, we calculate the expectation of $S_A$ as follows.
	\begin{align}
	\mathbb{E}\left\{S_A\right\} &= \mathbb{E}\left\{ \mathbb{E}\left\{S_A|v_{i-1} = d\right\} \right\} \nonumber\\&= \frac{\lambda}{2}\mathbb{E}\left\{d^2\right\} \nonumber\\&= \frac{\lambda}{2} \frac{2}{\mu^2} = \frac{\lambda}{\mu^2}. \nonumber
	\end{align}
	Thus, we have arrived at the conclusion in \eqref{11121600}.	
\end{IEEEproof}

\begin{lemma} \label{11121607}
	The area of sub-polygon $B$ can be obtained as
	\begin{align}\label{11121606}
	\mathbb{E}\left\{S_B\right\} = \frac{\lambda}{\mu^2}.
	\end{align}
\end{lemma}
\begin{IEEEproof}
	Sub-polygon $B$ is a rectangle, whose area is obtained as 
	\begin{align}
	S_B = N(v_{i-1}) v_i. \nonumber
	\end{align}
	Given the arrival rate of the original process, $N(v_{i-1})$ is independent with $v_i$. Thus, we have,
	\begin{align}
	\mathbb{E}\left\{S_B\right\} &= \mathbb{E}\left\{N(v_{i-1}) v_i\right\} \nonumber\\&= \mathbb{E}\left\{N(v_{i-1})\right\} \mathbb{E}\left\{ v_i\right\} \nonumber\\&= \frac{\lambda}{\mu} \frac{1}{\mu} = \frac{\lambda}{\mu^2}. \nonumber
	\end{align}
	Thus, we have arrived at the conclusion in \eqref{11121606}.
\end{IEEEproof}

\end{spacing}

\begin{spacing}{1.7}
\bibliographystyle{IEEEtran}
\bibliography{IEEEabrv,distortion}

\begin{thebibliography}{10}
\providecommand{\url}[1]{#1}
\csname url@samestyle\endcsname
\providecommand{\newblock}{\relax}
\providecommand{\bibinfo}[2]{#2}
\providecommand{\BIBentrySTDinterwordspacing}{\spaceskip=0pt\relax}
\providecommand{\BIBentryALTinterwordstretchfactor}{4}
\providecommand{\BIBentryALTinterwordspacing}{\spaceskip=\fontdimen2\font plus
\BIBentryALTinterwordstretchfactor\fontdimen3\font minus
  \fontdimen4\font\relax}
\providecommand{\BIBforeignlanguage}[2]{{%
\expandafter\ifx\csname l@#1\endcsname\relax
\typeout{** WARNING: IEEEtran.bst: No hyphenation pattern has been}%
\typeout{** loaded for the language `#1'. Using the pattern for}%
\typeout{** the default language instead.}%
\else
\language=\csname l@#1\endcsname
\fi
#2}}
\providecommand{\BIBdecl}{\relax}
\BIBdecl

\bibitem{diaz2016state}
M.~D{\'\i}az, C.~Mart{\'\i}n, and B.~Rubio, ``State-of-the-art, challenges, and
  open issues in the integration of internet of things and cloud computing,''
  \emph{Journal of Network and Computer Applications}, vol.~67, pp. 99--117,
  May 2016.

\bibitem{7845391}
G.~M. Dias, M.~Nurchis, and B.~Bellalta, ``Adapting sampling interval of sensor
  networks using on-line reinforcement learning,'' in \emph{IEEE 3rd World
  Forum on Internet of Things (WF-IoT)}, 2016.

\bibitem{li2013compressed}
S.~Li, L.~Da~Xu, and X.~Wang, ``Compressed sensing signal and data acquisition
  in wireless sensor networks and internet of things,'' \emph{IEEE Transactions
  on Industrial Informatics}, vol.~9, no.~4, pp. 2177--2186, November 2013.

\bibitem{4472240}
E.~J. Candes and M.~B. Wakin, ``An introduction to compressive sampling,''
  \emph{IEEE Signal Processing Magazine}, vol.~25, no.~2, pp. 21--30, March
  2008.

\bibitem{nyquist1928certain}
H.~Nyquist, ``Certain topics in telegraph transmission theory,''
  \emph{Transactions of the American Institute of Electrical Engineers},
  vol.~47, no.~2, pp. 617--644, April 1928.

\bibitem{1447892}
H.~J. Landau, ``Sampling, data transmission, and the nyquist rate,''
  \emph{Proceedings of the IEEE}, vol.~55, no.~10, pp. 1701--1706, October
  1967.

\bibitem{candes2006robust}
E.~J. Cand{\`e}s, J.~Romberg, and T.~Tao, ``Robust uncertainty principles:
  Exact signal reconstruction from highly incomplete frequency information,''
  \emph{IEEE Transactions on Information Theory}, vol.~52, no.~2, pp. 489--509,
  February 2006.

\bibitem{eldar2015sampling}
Y.~C. Eldar, \emph{Sampling theory: Beyond bandlimited systems}.\hskip 1em plus
  0.5em minus 0.4em\relax Cambridge University Press, 2015.

\bibitem{marvasti2012nonuniform}
F.~Marvasti, \emph{Nonuniform sampling: theory and practice}.\hskip 1em plus
  0.5em minus 0.4em\relax Springer Science \& Business Media, 2012.

\bibitem{vaswani2016recursive}
N.~Vaswani and J.~Zhan, ``Recursive recovery of sparse signal sequences from
  compressive measurements: A review.'' \emph{IEEE Transactions on Signal
  Processing}, vol.~64, no.~13, pp. 3523--3549, July 2016.

\bibitem{tropp2007signal}
J.~A. Tropp and A.~C. Gilbert, ``Signal recovery from random measurements via
  orthogonal matching pursuit,'' \emph{IEEE Transactions on Information
  Theory}, vol.~53, no.~12, pp. 4655--4666, December 2007.

\bibitem{maleki2010optimally}
A.~Maleki and D.~L. Donoho, ``Optimally tuned iterative reconstruction
  algorithms for compressed sensing,'' \emph{IEEE Journal of Selected Topics in
  Signal Processing}, vol.~4, no.~2, pp. 330--341, April 2010.

\bibitem{5419067}
M.~S. Asif and J.~Romberg, ``Dynamic updating for $\ell_{1}$minimization,''
  \emph{IEEE Journal of Selected Topics in Signal Processing}, vol.~4, no.~2,
  pp. 421--434, April 2010.

\bibitem{6827963}
------, ``Sparse recovery of streaming signals using $\ell_1$-homotopy,''
  \emph{IEEE Transactions on Signal Processing}, vol.~62, no.~16, pp.
  4209--4223, August 2014.

\bibitem{wijewardhana2017bayesian}
U.~L. Wijewardhana and M.~Codreanu, ``A bayesian approach for online recovery
  of streaming signals from compressive measurements,'' \emph{IEEE Transactions
  on Signal Processing}, vol.~65, no.~1, pp. 184--199, January 2017.

\bibitem{sun2017remote}
Y.~Sun, Y.~Polyanskiy, and E.~Uysal-Biyikoglu, ``Remote estimation of the
  wiener process over a channel with random delay,'' in \emph{IEEE
  International Symposium on Information Theory}, 2017.

\bibitem{bedewy2016optimizing}
A.~M. Bedewy, Y.~Sun, and N.~B. Shroff, ``Optimizing data freshness,
  throughput, and delay in multi-server information-update systems,'' in
  \emph{IEEE International Symposium on Information Theory}, 2016.

\bibitem{6195689}
S.~Kaul, R.~Yates, and M.~Gruteser, ``Real-time status: How often should one
  update?'' in \emph{IEEE International Conference on Computer Communications},
  2012.

\bibitem{6620189}
C.~Kam, S.~Kompella, and A.~Ephremides, ``Age of information under random
  updates,'' in \emph{IEEE International Symposium on Information Theory},
  2013.

\bibitem{7415972}
M.~Costa, M.~Codreanu, and A.~Ephremides, ``On the age of information in status
  update systems with packet management,'' \emph{IEEE Transactions on
  Information Theory}, vol.~62, no.~4, pp. 1897--1910, April 2016.

\bibitem{kam2016controlling}
C.~Kam, S.~Kompella, G.~D. Nguyen, J.~E. Wieselthier, and A.~Ephremides,
  ``Controlling the age of information: Buffer size, deadline, and packet
  replacement,'' in \emph{IEEE Military Communications Conference}, 2016.

\bibitem{kleinrock1975queueing}
L.~Kleinrock, \emph{Queueing system volume 1: Theory}.\hskip 1em plus 0.5em
  minus 0.4em\relax John Wiley \& Sons New York, 1975.

\bibitem{gross2008fundamentals}
D.~Gross, \emph{Fundamentals of queueing theory}.\hskip 1em plus 0.5em minus
  0.4em\relax John Wiley \& Sons New York, 2008.

\bibitem{ross1996stochastic}
S.~M. Ross, \emph{Stochastic processes}.\hskip 1em plus 0.5em minus 0.4em\relax
  John Wiley \& Sons New York, 1996.

\end{thebibliography}
\end{spacing}

\end{document}